# Respiratory Particle Deposition Probability due to Sedimentation with Variable Gravity and Electrostatic Forces

[1]Ioannis Haranas, [2]Ioannis Gkigkitzis, [3]George D. Zouganelis, [4]Maria K. Haranas, [2]Samantha Kirk

[1]Team Research Associate Department of Mathematics
East Carolina University
124 Austin Building, East Fifth Street Greenville
NC 27858-4353, USA
E-mail: yiannis.haranas@gmail.com

[2]Departments of Mathematics and Biomedical Physics,
East Carolina University
124 Austin Building, East Fifth Street Greenville
NC 27858-4353, USA
E-mail: gkigkitzisi@ecu.edu

[3]Health and Medical Sciences
Bournemouth and Poole College
North Road, Poole, Dorset,
BH14-OLS
United Kingdom
E-mail: zouganelisg@bpc.ac.uk

[4]Department of Informatics with Applications in Medicine
University of Thessaly
Argonafton &Filellinon, 38221 Volos, 74000
Magnesia, Greece
E-mail: maria_h1692@hotmail.com

## Abstract:

In this paper, we study the effects of the acceleration gravity on the sedimentation deposition probability, as well as the aerosol deposition rate on the surface of the Earth and Mars, but also aboard a spacecraft in orbit around Earth and Mars as well for particles with density $\rho_p = 1300$ kg/m$^3$, diameters $d_p = 1, 3, 5$ µm and residence times $t = 0.0272, 0.2$ s respectively. For particles of diameter 1µm we find that, on the surface of Earth and Mars the deposition probabilities are higher at the poles when compared to the ones at the equator. Similarly, on the surface of the Earth we find that the deposition probabilities exhibit 0.5% and 0.4 % higher percentage difference at the poles when compared to that of the equator, for the corresponding residence times. Moreover in orbit equatorial orbits result to higher deposition probabilities when compared to polar ones. For both residence times particles with the diameters considered above in circular and elliptical orbits around Mars, the deposition probabilities appear to be the same for all orbital inclinations. Sedimentation probability increases drastically with particle diameter and orbital eccentricity of the orbiting spacecraft. Finally, as an alternative framework for the study of interaction and the effect of gravity in biology, and in particular gravity and the respiratory system we introduce is the term information in a way Shannon has introduced it, considering the sedimentation probability as a random variable. This can be thought as a way in which gravity enters the cognitive processes of the system (processing of information) in the cybernetic sense.

**1.Introduction**

Airborne particles have always been around in the history of mankind. Human population has been exposed to the inhalation of airborne particle clouds, clouds that have been created by various natural processes that include volcanic eruptions and forest fires. The severity of corresponding exposures can be hazardous or even fatal, if the particle air concentration, size, and composition of the suspended materials. In a pioneering paper by Watkins Pitchford and Moir (1916) the authors understand the importance of that, and they find that 80% of the particles in silicotic human lungs are smaller that 2µm in diameter. Next, Hatch and Gross (1964) report that inhalation exposure to aerosols of fine zinc oxide particles of diameter less 0.6µm can cause the so called "metal fume fever", where exposure to similar air concentrations of zinc oxide that results from bulk material is not toxic. The measured concentration of the corresponding contaminant particles and their total inhaled volume was found to be insufficient to predict the pathological effects of inhalation exposure, and also to predict the relative hazard to a particular aerosol, that might be dust or mist. It is the size distribution and also the density of particular aerosol particles which determines their penetration depth as well as their fractional deposition in the respiratory track which results in the determination of the location of their critical sites of action and finally their translocational mechanisms. The use of the term aerosol simply refers to particles that might be solid or liquid in nature, with a sufficiently small diameter that can be suspended in the air as it is defined by Green and Lane (1957).

To describe any possible risks associated with the inhalation of aerosol particles, one must know how much is deposited in a particular region of the respiratory track as well as how much is the left after a physiological clearance from that region. Any remaining material can constitute a potential effective dose which can produce an acute or a chronic pulmonary disease. Deposition of the aerosol material within the respiratory track is predicted by models and also experimental evaluation and they important in the case of soluble aerosols, because their contaminants reach the bloodstream but also the lymphatic channels from several places of the respiratory track. Deposition is an important step in determining various clearance processes. In Morrow et al. (1966) a discussion on lung dynamics the respiratory track can be described in terms of three compartments that are based upon a clearance mechanism associated with each place separately. Inhaled particles deposited in the posterior nares will be caught in the mucus and conveyed by mucocilliary action through the nasopharynx, and into the gastrointestinal track (Holmes et al., 1950)

In this contribution we examine the effect of variable gravity might have on the deposition probability of aerosol particles on the surface of the Earth and also above an orbiting spacecraft. For that we correct the acceleration of gravity on the surface for the Earth's rotation and for its oblateness via the $J_2$ harmonic coefficient of the gravitational field. The effect of various latitudes is examined and compared, for particles of various sizes and a particular density. Similarly, in orbiting spacecraft this is achieved by transforming the gravitational acceleration at the orbital altitude of the spacecraft as a function of the orbital elements, and for three orbits of three different inclinations the sedimentation probabilities are examined and compared. Next, we modify the acceleration of gravity by adding an electrostatic deposition acceleration caused by the image force acting on the particles. This acceleration is responsible for the deposition increase within the respiratory track, and therefore the deposition probability is again calculated on the surface of the Earth and in an orbiting spacecraft experiment. After that, the aerosol deposition rate on the surface of the Earth and Mars as well as aboard an orbiting spacecraft is also calculated. Finally, as an alternative framework for the study of interaction of the effect of gravity in biology, and in particular gravity and respiratory system we introduce, is the term information in a way Shannon has introduced it, considering the sedimentation probability as a random variable. This can be thought as a way in which gravity enters the cognitive processes of the system (processing of information) in the cybernetic sense.

## 2. Mechanisms of Deposition

Aerosol particles are inhaled and deposited in the human respiratory track. Inertial impaction of the inhaled particles is the main mechanism of large particle deposition in the upper part of the respiratory track. Inertial impaction can act on particles ranging from 2 to 3 μm to greater than 20 μm in diameter, i.e. 50 μm. It is the inertia of the large airborne particles that tent to maintain their initial path when the supporting air stream is deflected suddenly by nasal turbinates or branching of the airways. The probability of inertial deposition $P_{dep_i}$ is proportional to the product of the terminal settling velocity $v_{ter}$, which is higher for larger diameter particles, and $v_{air}$, is the entrained particle velocity which is inversely proportional to the radius $R$ of the airway (Gussman, 1965a):

$$P_{dep_i} \propto \frac{v_{ter} v_{air} \sin \theta}{(gR + v_{ter} v_{air} \sin \theta)}, \tag{1}$$

where $g$ is the acceleration of gravity. In the case of a large increasing air velocity, smaller airway radius and large bending angle $\theta$ implies a large inertial impaction probability $P_i$ becomes ideally large.

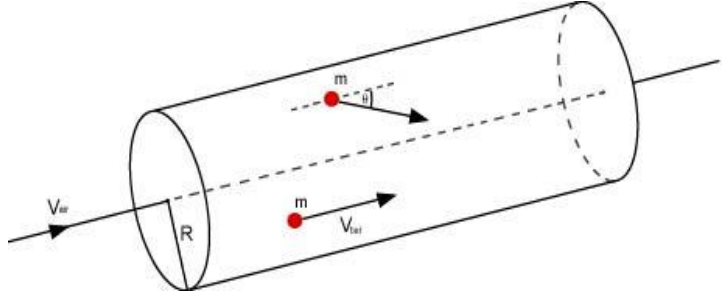

**Fig. 1** Particles entering the airway at an angle $\theta$ with terminal velocity $v_{ter}$ and air velocity $v_{air}$, at an angle $\theta$ to the horizontal

Settling terminal velocity is related to sedimentation, and it is very important for particle deposition in the respiratory track. A particle falling under gravity, acquires a terminal velocity that is given by (Haranas et al., 2012):

$$v_{ter} = \frac{g d^2 \left( \rho_p - \rho_{air} \right)}{18 \eta_{air}},$$ (2)

where $\eta_{air}$ is the viscosity of the air, $\rho_p$ is the density of the particle, $\rho_{air}$ is the density of the air, and $d$ is the diameter of the particle. When the particle's diameter becomes very small and of the same order with the air molecule mean free path $L$, the corresponding drag increases and the following correction bust be applied (Cunningham 1910):

$$v_{ter}\left(cor\right) = v_{ter}\left(cal\right)\left[1 + \frac{2C_c}{d}L\right],$$ (3)

where $v_{ter}\left(cor\right)$ and $v_{ter}\left(cal\right)$ are the corrected and calculated terminal velocities respectively (Gussman, 1969), and where $C_c$ is given by the following expression (Hinds 1999):

$$C_c = 1 + \frac{\lambda}{d_p}\left[2.34 + 1.05e^{-\left(\frac{0.39}{\lambda}d_p\right)}\right], \qquad (4)$$

$C_c$ is the so called Cunningham correction or "slip correction factor" that must be taken intro account when the size of the aerosol particles is comparable to the mean free path of the atmospheric molecules, discontinuities in the in the medium become important consideration, and particle mobility increases. $\lambda$ is the mean free path of the air, and its inversely proportional to its density $\lambda = 7.91 \times 10^{-8} / \rho_{air}$ for $\lambda$ in meters and $\rho_{air}$ in kg/m$^3$, and $d_p$ in meters. In a time $t$, the inhaled particles travel a distance $d_{ter} = v_{ter}t$, and therefore, if the particles enter the airway at an angle $\theta$ with the horizontal then the deposition probability $P_{dep}$ due to sedimentation becomes (Landhal, 1950):

$$P_{sed} = 1 - e^{-\left(\frac{\tau(\rho_p - \rho_{air})d_p^2 C_c}{18\eta_{air}R}g\cos\theta\right)}. \qquad (5)$$

The time it takes for a particle to travel in the trachea is:

$$\tau = \frac{L_{tr}}{v_{air}}, \qquad (6)$$

eliminating this traveling time $\tau$ in equation (5) we obtain that the sedimentation probability it takes the form:

$$P_{sed} = 1 - e^{-\left(\frac{(\rho_p - \rho_{air})d_p^2 C_c}{18\eta_{air}R}\frac{L}{v_{air}}g\cos\theta\right)} \qquad (7)$$

where $L$ is the length of trachea, and $v$ the velocity of the air, and $R$ the radius of the airway. Sedimentation is the basic mechanism via which inhaled particles in the range $0.1\,\mu\text{m} \leq d_p \leq 50\,\mu\text{m}$ are deposited (Morrow, 1964). We can write the particle's effective distance at right angles to the direction of travel in a following way:

$$h_s = \frac{v_{ter}\eta_{air}}{g}\sin\theta , \qquad (8)$$

where $\theta$ is the same angle define before. Finally, the ratio $r$ of the fall distance or the distance that the particle travels in time t moving with terminal velocity $v_t$ to the max distance of deposition or the physical distance of the diameter of object (trachea) in our case that the particle has to travel becomes:

$$h_{max} = \frac{v_{ter}\tau}{2R}\cos\theta = \frac{v_{ter}}{2R}\left(\frac{L_{tr}}{v_{air}}\right)\cos\theta . \qquad (9)$$

When a particle enters our respiratory system is subject to the deposition mechanism described above. The actual deposition efficiency of a given particle size has been determined experimentally. Various models have been developed to predict the deposition based on experimental data. The most advanced and widely used is the one developed by the International Commission on Radiological Protection (ICRP) and the National Council on Radiation Protection and Measurement (NCRP). The total deposition fraction (DF) in the respiratory system according to ICRP model is given by (ICRP, 1994):

$$DF = \left[1 - 0.5\left[1 - \frac{1}{1 + 0.00076 d_p^{2.8}}\right]\right]\left[0.0587 + \frac{0.911}{1 + e^{(4.77 + 1.485\ln d_p)}} + \frac{0.943}{1 + e^{(0.53 - 2.58\ln d_p)}}\right]$$
. $\qquad (10)$

where $d_p$ is the particle diameter in microns.

## 3. The gravitational acceleration on the surface of the Earth and at orbital point

In our effort to study the effect of gravity on the respiratory system, let us consider the acceleration of gravity $g$ at the orbital altitude of the spacecraft in which a respirator type of experiment is taking place under con-trolled conditions. Following Haranas, et al., (2013) we write the acceleration of gravity as the sum of three different components namely:

$$V_{tot}(r') = -\frac{GM_E}{r'} + \frac{GM_E R_E^2 J_2}{2r'^3}\left(3\sin^2\phi_E - 1\right) - \frac{1}{2}\omega_E^2 r'^2 \cos^2\phi_E. \quad (11)$$

where $r'$ is the radial distance from the center of the Earth to an external surface point, $M_E$ is the mass of the Earth, $R_E$ is the radius of the Earth, $J_2$ is the zonal harmonic coefficient that describes the oblateness of the Earth, $\omega_E$ the angular velocity of the Earth, and $\phi_E$ is the geocentric latitude of the designed experiment. Zonal harmonics are simply bands of latitude, whose boundaries are the roots of a Legendre polynomial. Spherical harmonics in general are very important concept in solar system research and in particular in the modelling of planetary gravity fields. For example in Hadjifotinou (2000) the author uses a gravitational potential that includes a $J_2$ as well as a $J_4$ harmonics in predicting numerically the motion of Saturn's satellites. Similarly, in Iorio (2011) the author derives the precession of the ascending node of a satellite due to Lense-Thirring effect as a function of the $J_2$ harmonic. This particular gravitational harmonic coefficient is a result of the Earth's shape and is about 1000 times larger than the next harmonic coefficient $J_3$ and its value is equal to $J_2 = -0.0010826260$ (Kaula, 2000). At the orbital point of the spacecraft the rotational potential on the surface of the earth does not affect the orbit of spacecraft. Therefore the gravitational acceleration as it is given by Equ. (2) can be transformed as a function of orbital elements, using standard transformations given by Kaula (2000) and Vallado (2007) namely $\sin\phi_E = \sin i \sin(u) = \cos\theta_E$, where, $\phi_E$ is the geocentric latitude, measured from the Earth's equator to the poles, and $\theta_E$ is the corresponding colatitude measured from the poles down to the equator ($\theta_E = 90 - \phi_E$), $u = \omega + f$ is the argument of latitude that defines the position of a body moving along a Kepler orbit, $i$ its orbital inclination, $\omega$ is the argument of the perigee of the spacecraft (not to be confused with angular velocity, which we write with subscripts – see nomenclature section below), $f$ is its true anomaly (an angle defined between the orbital position of the spacecraft and its perigee). The orbital elements are shown in figure 2 below.

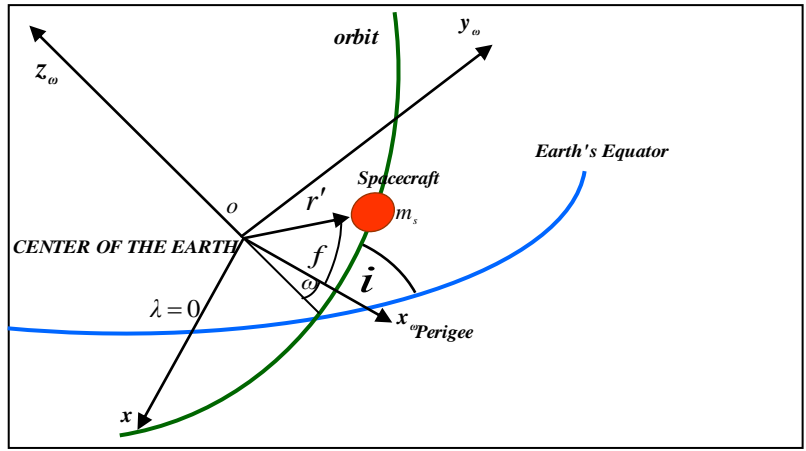

**Figure. 1** Explanation of the orbital elements: inclination $i$, argument of latitude $u = \omega + f$ , and the radial vector $r'$ of the spacecraft, and $\lambda = 0$ is the zero longitude point on the Earth's equator, and $x_\omega, y_\omega, z_\omega$ define a right handed coordinate system (Haranas et al., 2013).

To familiarize the reader with the orbital elements used here let us define the orbital elements appearing in our Eq. (4) below. $a_s$ is the semi-major axis that defines the size of the orbital ellipse. It is the distance from the center of the ellipse to an apsis i.e. the point where the radius vector is maximum or minimum (i.e. apogee and perigee points). Similarly, $e$ is the eccentricity, that defines the shape of the orbital ellipse (minor to major axis ratio), $i$ is the inclination of the orbit defined as the angle between the orbital and equatorial planes, and $\omega$ is the argument of the perigee, the angle between the direction of the ascending node (the point on the equatori-al plane at which the spacecraft crosses from south to north) and the direction of the perigee. Finally, the true anomaly $f$, is the angle that locates the spacecraft in the orbital ellipse and is measured in the direction of motion from the perigee to the position vector of the satellite. Assuming an elliptical orbit the geocentric orbital distance $r'$ is given by (Vallado, 2007):

$$r' = \frac{a_s \left(1 - e^2\right)}{\left(1 + e \cos f\right)} \ , \tag{12}$$

and therefore Eq. (11) becomes a function of the spacecraft orbital elements:

$$g_{tot} = -\frac{GM_E \left(1 + e \cos f\right)^2}{a_s^2 \left(1 - e^2\right)^2} + \frac{3 GM_E R_E^2 J_2 \left(1 + e \cos f\right)^4}{2 a_s^4 \left(1 - e^2\right)^4} \left(3 \sin^2 i \sin^2 f - 1\right).$$

$$\tag{13}$$

First, let us consider an sedimentation experiment that takes place on the surface of the Earth $r = R_E$. In this case from Eq. (11) we obtain the following expression for the acceleration of gravity corrected for the oblateness and the rotation of the Earth on the Earth's surface and therefore the gravitational acceleration $g_s$ becomes:

$$- g_s = \frac{GM_E}{R_E^2} - \frac{3 GM_E J_2}{2 R_E^2} \left(3 \sin^2 \phi - 1\right) - R_E \omega_E^2 \cos^2 \phi \ . \tag{14}$$

Furthermore, thinking the Earth to be an ellipsoid of revolution we can write that (Kaula, 2000):

$$R_E = R_{eq} \left(1 - \left(f' + \frac{3}{2} f'^2\right) \sin^2 \phi_E + \frac{3}{2} f'^2 \sin^4 \phi_E - \dots\right) \approx R_{eq} \left(1 - f' \sin^2 \phi_E\right), \tag{15}$$

where $\varphi$ is the planetocentric latitude, and $f'$ is the Earth's flattering that it is given by:

$$f' = \frac{R_{eq} - R_{pol}}{R_{eq}} = \frac{3 J_2}{2} + \frac{R_E^3 \omega_E^2}{2 G M_E} \ , \tag{16}$$

where $R_{eq}$ and $R_{pol}$ is the Earth's equatorial and polar radii and therefore. Eq. (14) becomes:

$$g_s = \frac{GM_E}{R_{eq}^2\left(1 - f'\sin^2\phi\right)^2}$$
$$- \frac{3GM_E J_2}{2R_{eq}^2\left(1 - f'\sin^2\phi\right)^2}\left(3\sin^2\phi - 1\right) - R_{eq}\left(1 - f'\sin^2\phi\right)\omega_E^2\cos^2\phi \quad .(17)$$

On the surface of the Earth and for the geocentric latitude $\phi = 0^\circ$, 45°, 90° Eq. (17) results in the following expressions for the gravitational acceleration:

$$g_{\varphi=0} = \frac{GM_E}{R_{eq}^2} + \frac{3GM_E J_2}{2R_{eq}^2} - R_{eq}\omega_E^2 \quad . \tag{18}$$

$$g_{\varphi=45} = \frac{GM_E}{R_{eq}^2\left(1 - f'\right)^2} - \frac{3GM_E J_2}{4R_{eq}^2\left(1 - \dfrac{f'}{2}\right)^2} - \frac{1}{2}\left(1 - \frac{f'}{2}\right)R_{eq}\omega_E^2 \quad , \tag{19}$$

$$g_{\varphi=90} = \frac{GM_E}{R_{eq}^2\left(1 - f'\right)^2} - \frac{3GM_E J_2}{R_{eq}^2\left(1 - f'\right)^2} \quad . \tag{20}$$

## 4. Modified gravity probability deposition due to sedimentation

First let us now consider Eq. (7) for the deposition probability of particles on the surface of the Earth, where the gravitational acceleration has been corrected for the rotation of the Earth and also for its oblateness via the $J_2$ harmonic coefficient. Using Eqs (6), (18), (19), (20) we can write that the deposition probability becomes:

$$P_{sed} = 1 - e^{-\left\{\frac{(\rho_p - \rho_{air})d_p^2 C_c}{18\eta_{air}R}\frac{L}{v_{air}}\left(\frac{GM_E}{R_{eq}^2} + \frac{3GM_E J_2}{2R_{eq}^2} - R_{eq}\omega_E^2\right)\cos\theta\right\}}, \tag{21}$$

$$P_{sed} = 1 - e^{-\left(\frac{(\rho_p - \rho_{air})d_p^2 C_c}{18\eta_{air}R}\frac{L}{v_{air}}\left(\frac{GM_E}{R_{eq}^2\left(1-f'\right)^2} - \frac{3GM_E J_2}{4R_{eq}^2\left(1-\frac{f'}{2}\right)^2} - \frac{1}{2}\left(1-\frac{f'}{2}\right)R_{eq}\omega_E^2\right)\cos\theta\right)}, \qquad (22)$$

$$P_{sed} = 1 - e^{-\left(\frac{(\rho_p - \rho_{air})d_p^2 C_c}{18\eta_{air}R}\frac{L}{v_{air}}\left(\frac{GM_E}{R_{eq}^2\left(1-f'\right)^2} - \frac{3GM_E J_2}{R_{eq}^2\left(1-f'\right)^2}\right)\cos\theta\right)}. \qquad (23)$$

For circular orbits i.e. $e = 0$, of inclinations $i = 0°, 45°, 90°$ we obtain:

$$P_{sed} = 1 - e^{-\left(\frac{(\rho_p - \rho_{air})d_p^2 C_c}{18\eta_{air}R}\frac{L}{v_{air}}\left(\frac{GM_E}{a^2} + \frac{3GM_E R_{eq}^2 J_2}{2a^4}\right)\cos\theta\right)} \qquad (24)$$

$$P_{sed} = 1 - e^{-\left(\frac{(\rho_p - \rho_{air})d_p^2 C_c}{18\eta_{air}R}\frac{L}{v_{air}}\left(\frac{GM_E}{a^2} + \frac{3GM_E R_{eq}^2 J_2}{2a^4}\left(\frac{3\sin^2 u}{2} - 1\right)\right)\cos\theta\right)}, \qquad (25)$$

$$P_{sed} = 1 - e^{-\left(\frac{(\rho_p - \rho_{air})d_p^2 C_c}{18\eta_{air}R}\frac{L}{v_{air}}\left(\frac{GM_E}{a^2} + \frac{3GM_E R_{eq}^2 J_2}{2a^4}\left(\frac{3\sin^2 u - 1}{2}\right)\right)\cos\theta\right)}. \qquad (26)$$

Similarly, for elliptical orbits of the same inclinations the deposition probability of particles takes the form:

$$P_{sed} = 1 - e^{-\left(\frac{(\rho_p - \rho_{air})d_p^2 C_c}{18\eta_{air}R}\frac{L}{v_{air}}\left(\frac{GM_E\left(1+e\cos f\right)^2}{a^2\left(1-e^2\right)} + \frac{3GM_E R_{eq}^2 J_2\left(1+e\cos f\right)^4}{2a^4\left(1-e^2\right)^4}\right)\cos\theta\right)} \qquad (27)$$

$$P_{sed} = 1 - e^{-\left(\frac{(\rho_p - \rho_{air})d_p^2 C_c}{18\eta_{air}R}\frac{L}{v_{air}}\left(\frac{GM_E\left(1+e\cos f\right)^2}{a^2\left(1-e^2\right)} + \frac{3GM_E R_{eq}^2 J_2\left(1+e\cos f\right)^4}{2a^4\left(1-e^2\right)^4}\left(\frac{3\sin^2 f}{2} - 1\right)\right)\cos\theta\right)}$$

$$\qquad (28)$$

$$P_{sed} = 1 - e^{-\left(\frac{(\rho_p - \rho_{air})d_p^2 C_c}{18\eta_{air}R}\frac{L}{v_{air}}\left(\frac{GM_E\left(1+e\cos f\right)^2}{a^2\left(1-e^2\right)} + \frac{3GM_E R_{eq}^2 J_2\left(1+e\cos f\right)^4}{2a^4\left(1-e^2\right)^4}\left(\frac{3\sin^2 f - 1}{2}\right)\right)\cos\theta\right)}. \qquad (29)$$

## 5. Modified gravity ratio of the fall distance to the maximum distance for deposition

In Eq. (7) we have given the ratio $r$ of the fall distance to the max distance for deposition. In this section we correct equation (7) by correcting the gravitational acceleration on the surface of the Earth for the rotation of the Earth, and also for its oblateness, we give expressions for three different geocentric latitudes, taking Earth's flattering into account. Similarly, we extend the expression for the maximum height when in orbit around a planetary body in our case the Earth and therefore we obtain:

$$h_{max} = \frac{\left(\rho_p - \rho_{air}\right) d_p^2 C_c}{18 \eta_{air} R} \frac{L}{v_{air}} \left( \frac{GM_E}{R_{eq}^2} + \frac{3GM_E J_2}{2R_{eq}^2} - R_{eq} \omega_E^2 \right) \cos \theta \,, \qquad (30)$$

$$h_{max} = \frac{\left(\rho_p - \rho_{air}\right) d_p^2 C_c}{18 \eta_{air} R} \frac{L}{v_{air}} \left( \frac{GM_E}{R_{eq}^2 \left(1 - f'\right)^2} - \frac{3GM_E J_2}{4R_{eq}^2 \left(1 - \frac{f'}{2}\right)^2} - \frac{1}{2}\left(1 - \frac{f'}{2}\right) R_{eq} \omega_E^2 \right) \cos \theta \,,$$

$$(31)$$

$$h_{max} = \frac{\left(\rho_p - \rho_{air}\right) d_p^2 C_c}{18 \eta_{air} R} \frac{L}{v_{air}} \left( \frac{GM_E}{R_{eq}^2 \left(1 - f'\right)^2} - \frac{3GM_E J_2}{R_{eq}^2 \left(1 - f'\right)^2} \right) \cos \theta \,. \qquad (32)$$

For circular orbits i.e. $e = 0$, of inclinations $i = 0°$, $45°$, $90°$ we obtain:

$$h_{max} = \frac{\left(\rho_p - \rho_{air}\right) d_p^2 C_c}{18 \eta_{air} R} \frac{L}{v_{air}} \left( \frac{GM_E}{a^2} + \frac{3GM_E R_{eq}^2 J_2}{2a^4} \right) \cos \theta \,, \qquad (33)$$

$$h_{max} = \frac{\left(\rho_p - \rho_{air}\right) d_p^2 C_c}{18 \eta_{air} R} \frac{L}{v_{air}} \left( \frac{GM_E}{a^2} + \frac{3GM_E R_{eq}^2 J_2}{2a^4} \left( \frac{3\sin^2 u}{2} - 1 \right) \right) \cos \theta \,,$$

$$(34)$$

$$h_{max} = \frac{\left(\rho_p - \rho_{air}\right) d_p^2 C_c}{18 \eta_{air} R} \frac{L}{v_{air}} \left( \frac{GM_E}{a^2} + \frac{3GM_E R_{eq}^2 J_2}{2a^4} \left( \frac{3\sin^2 u - 1}{2} \right) \right) \cos \theta \,.$$

$$(35)$$

Similarly, for elliptical orbits of the same inclinations the ratio takes the form:

$$h_{max} = \frac{(\rho_p - \rho_{air})d_p^2 C_c}{18\eta_{air}R} \frac{L}{v_{air}} \left( \frac{GM_E(1+e\cos f)^2}{a^2(1-e^2)} + \frac{3GM_E R_{eq}^2 J_2(1+e\cos f)^4}{2a^4(1-e^2)^4} \right) \cos\theta$$

,
$$(36)$$

$$h_{max} = \frac{(\rho_p - \rho_{air})d_p^2 C_c}{18\eta_{air}R} \frac{L}{v_{air}} \left( \frac{GM_E(1+e\cos f)^2}{a^2(1-e^2)} + \frac{3GM_E R_{eq}^2 J_2(1+e\cos f)^4}{2a^4(1-e^2)^4} \left( \frac{3\sin^2 f}{2} - 1 \right) \right) \cos\theta$$

,
$$(37)$$

$$h_{max} = \frac{(\rho_p - \rho_{air})d_p^2 C_c}{18\eta_{air}R} \frac{L}{v_{air}} \left( \frac{GM_E(1+e\cos f)^2}{a^2(1-e^2)} + \frac{3GM_E R_{eq}^2 J_2(1+e\cos f)^4}{2a^4(1-e^2)^4} \left( \frac{3\sin^2 f - 1}{2} \right) \right) \cos\theta$$

.
$$(38)$$

## 6. The Inclusion of Electrostatic Deposition Forces

Up to equation (38) we have dealt with the effect of a variable gravity on the surface of a planetary body and also in an experiment above an orbiting spacecraft. Next, we will consider dust particle electrostatic forces first in the absence of an electric field $\vec{E}$ falling in the trachea. Following Yu (1985) we say that there are two electric forces which cause particles to deposit on the surface of an enclosure. One is the image force due to the interaction between a particle and the wall. The other is the space charge force due to the mutual repulsion between particles of the same charge. The image force is a single particle effect, while the space charge force depends on the particle concentration of the system. In order to find out which of these two forces it's important in lung deposition, the inter-particle distance must be estimated in relation to the airways. Because most environmental aerosols have low concentration numbers, we will assume that the primary deposition force is the image force acting on the particles, which is also responsible for the deposition increase within the respiratory track. Because the particle is acted upon by the vector sum of many forces, for the determination of an exact image deposition acceleration will be a very hard task. We simplify the analysis by following the idea give in Yu (1985) where the total acceleration acted upon the particle is a superposition of the gravitational and that resulted by the image force, and therefore we have that:

$$a_{tot} = g\cos\theta + a_{image}\cos\xi \ , \tag{39}$$

where $a_{image} \ \xi$ is the vertical component of $a_{image,}$, where $\xi$ is the angle between the vertical component of the image acceleration and the direction of the acceleration of gravity. Therefore eq. (1) can be modified to:

$$P_i \propto \frac{v_{ter}v_{air}\sin\theta}{\left(\left(g + a_{image}\cos\xi\right)R + v_{ter}v_{air}\sin\theta\right)} \quad . \tag{40}$$

Therefore using the expression for the image force given by Yu (1985) we obtain the vertical component of the image acceleration to:

$$a_{image} = \frac{F_{image}}{m} = \frac{q^2 r^2 \cos\xi}{16\pi\varepsilon_0 m(R-r)^2} \ , \tag{41}$$

where $R$ is the tracheal radius, and $r$ is the distance of the particle from the axis, $m$ is the mass of the particle, and $q$ is its charge, and $\varepsilon_0$ is the permittivity of the air. Therefore the sedimentation probability in Eq. (7) takes the form:

$$P_{sed} = 1 - e^{-\left(\frac{\left(\rho_p - \rho_{air}\right)d_p^2 C_c}{18\eta_{air}R} \frac{L}{v_{air}}\left(g\cos\theta + a_{image}\cos\xi\right)\right)} \quad . \tag{42}$$

Next, we can write a particle effective distance at right angles to the direction of travel to be:

$$h_s = \frac{v_{ter}\eta_{air}}{\left(g + a_{image}\cos\xi\right)}\sin\theta \ . \tag{43}$$

First let us now consider Eq. (1) for the deposition probability of particles on the surface of the Earth, where the gravitational acceleration has been corrected for the rotation of the Earth and also for its oblateness via the $J_2$ harmonic coefficient. Using Eqs (9), (10), (11) we can write that:

$$P_{sed} = 1 - e^{-\left(\frac{(\rho_p - \rho_{air})d_p^2 C_c}{18\eta_{air}R}\frac{L}{v_{air}}\left(\left(\frac{GM_E}{R_{eq}^2} + \frac{3GM_E J_2}{2R_{eq}^2} - R_{eq}\omega_E^2\right)\cos\theta + \frac{q^2 r^2 \cos\xi}{16\pi\varepsilon_0 m(R-r)^2}\right)\right)}, \qquad (44)$$

$$P_{sed} = 1 - e^{-\left(\frac{(\rho_p - \rho_{air})d_p^2 C_c}{18\eta_{air}R}\frac{L}{v_{air}}\left(\left(\frac{GM_E}{R_{eq}^2(1-f')} + \frac{3GM_E J_2}{4R_{eq}^2\left(1-\frac{f'}{2}\right)^2} - \frac{1}{2}\left(1-\frac{f'}{2}\right)R_{eq}\omega_E^2\right)\cos\theta + \frac{q^2 r^2 \cos\xi}{16\pi\varepsilon_0 m(R-r)^2}\right)\right)}, (45)$$

$$P_{sed} = 1 - e^{-\left(\frac{(\rho_p - \rho_{air})d_p^2 C_c}{18\eta_{air}R}\frac{L}{v_{air}}\left(\left(\frac{GM_E}{R_{eq}^2(1-f')^2} + \frac{3GM_E J_2}{R_{eq}^2(1-f')^2}\right)\cos\theta + \frac{q^2 r^2 \cos\xi}{16\pi\varepsilon_0 m(R-r)^2}\right)\right)}. \qquad (46)$$

For circular orbits i.e. $e = 0$, of inclinations $i = 0°$, 45°, 90° we obtain:

$$P_{sed} = 1 - e^{-\left(\frac{(\rho_p - \rho_{air})d_p^2 C_c}{18\eta_{air}R}\frac{L}{v_{air}}\left(\left(\frac{GM_E}{a^2} + \frac{3GM_E R_{eq}^2 J_2}{2a^4}\right)\cos\theta + \frac{q^2 r^2 \cos\xi}{16\pi\varepsilon_0 m(R-r)^2}\right)\right)}, \qquad (47)$$

$$P_{sed} = 1 - e^{-\left(\frac{(\rho_p - \rho_{air})d_p^2 C_c}{18\eta_{air}R}\frac{L}{v_{air}}\left(\left(\frac{GM_E}{a^2} + \frac{3GM_E R_{eq}^2 J_2}{2a^4}\left(\frac{3\sin^2 u}{2} - 1\right)\right)\cos\theta + \frac{q^2 r^2 \cos\xi}{16\pi\varepsilon_0 m(R-r)^2}\right)\right)}, (48)$$

$$P_{sed} = 1 - e^{-\left(\frac{(\rho_p - \rho_{air})d_p^2 C_c}{18\eta_{air}R}\frac{L}{v_{air}}\left(\left(\frac{GM_E}{a^2} + \frac{3GM_E R_{eq}^2 J_2}{2a^4}\left(\frac{3\sin^2 u - 1}{2}\right)\right)\cos\theta + \frac{q^2 r^2 \cos\xi}{16\pi\varepsilon_0 m(R-r)^2}\right)\right)}. (49)$$

Similarly, for elliptical orbits of the same inclinations the particle deposition probability takes the form:

$$P_{sed} = 1 - e^{-\left(\frac{(\rho_p - \rho_{air})d_p^2 C_c}{18\eta_{air}R}\frac{L}{v_{air}}\left(\left(\frac{GM_E(1+e\cos f)^2}{a^2(1-e^2)} + \frac{3GM_E R_{eq}^2 J_2(1+e\cos f)^4}{2a^4\left(1-e^2\right)^4}\right)\cos\theta + \frac{q^2 r^2 \cos\xi}{16\pi\varepsilon_0 m(R-r)^2}\right)\right)}, (50)$$

$$P_{sed} = 1 - e^{-\left(\frac{(\rho_p - \rho_{air})d_p^2 C_c}{18\eta_{air}R}\frac{L}{v_{air}}\left(\left(\frac{GM_E(1+e\cos f)^2}{a^2(1-e^2)} + \frac{3GM_E R_{eq}^2 J_2(1+e\cos f)^4}{2a^4\left(1-e^2\right)^4}\left(\frac{3\sin^2 f}{2} - 1\right)\right)\cos\theta + \frac{q^2 r^2 \cos\xi}{16\pi\varepsilon_0 m(R-r)^2}\right)\right)}$$

$$(51)$$

$$P_{sed} = 1 - e^{-\left(\frac{(\rho_p - \rho_{air})d_p^2 C_c}{18\eta_{air}R}\frac{L}{v_{air}}\left(\left(\frac{GM_E(1+e\cos f)^2}{a^2(1-e^2)} + \frac{3GM_E R_{eq}^2 J_2(1+e\cos f)^4}{2a^4\left(1-e^2\right)^4}\left(\frac{3\sin^2 f - 1}{2}\right)\right)\cos\theta + \frac{q^2 r^2 \cos\xi}{16\pi\varepsilon_0 m(R-r)^2}\right)\right)}. (52)$$

Similarly, we extend the expression for the maximum height when in orbit around a planetary body in our case the Earth and therefore we obtain:

$$h_{max} = \frac{(\rho_p - \rho_{air}) d_p^2 C_c}{18 \eta_{air} R} \frac{L}{v_{air}} \left[ \left( \frac{GM_E}{R_{eq}^2} + \frac{3GM_E J_2}{2R_{eq}^2} - R_{eq} \omega_E^2 \right) \cos\theta + \frac{q^2 r^2 \cos\xi}{16\pi\varepsilon_0 m(R-r)^2} \right] \quad (53)$$

$$h_{max} = \frac{(\rho_p - \rho_{air}) d_p^2 C_c}{18 \eta_{air} R} \frac{L}{v_{air}} \left[ \left( \frac{GM_E}{R_{eq}^2 (1-f')^2} + \frac{3GM_E J_2}{4R_{eq}^2 \left(1-\frac{f'}{2}\right)^2} - \frac{1}{2}\left(1-\frac{f'}{2}\right) R_{eq} \omega_E^2 \right) \cos\theta + \frac{q^2 r^2 \cos\xi}{16\pi\varepsilon_0 m(R-r)^2} \right] , (54)$$

$$h_{max} = \frac{(\rho_p - \rho_{air}) d_p^2 C_c}{18 \eta_{air} R} \frac{L}{v_{air}} \left[ \left( \frac{GM_E}{R_{eq}^2 (1-f')^2} + \frac{3GM_E J_2}{R_{eq}^2 (1-f')^2} \right) \cos\theta + \frac{q^2 r^2 \cos\xi}{16\pi\varepsilon_0 m(R-r)^2} \right] . \quad (55)$$

For circular orbits i.e. $e = 0$, of inclinations $i = 0°$, $45°$, $90°$ we obtain:

$$h_{max} = \frac{(\rho_p - \rho_{air}) d_p^2 C_c}{18 \eta_{air} R} \frac{L}{v_{air}} \left[ \left( \frac{GM_E}{a^2} + \frac{3GM_E R_{eq}^2 J_2}{2a^4} \right) \cos\theta + \frac{q^2 r^2 \cos\xi}{16\pi\varepsilon_0 m(R-r)^2} \right], \quad (56)$$

$$h_{max} = \frac{(\rho_p - \rho_{air}) d_p^2 C_c}{18 \eta_{air} R} \frac{L}{v_{air}} \left[ \left( \frac{GM_E}{a^2} + \frac{3GM_E R_{eq}^2 J_2}{2a^4}\left(\frac{3\sin^2 u}{2}-1\right) \right) \cos\theta + \frac{q^2 r^2 \cos\xi}{16\pi\varepsilon_0 m(R-r)^2} \right], (57)$$

$$h_{max} = \frac{(\rho_p - \rho_{air}) d_p^2 C_c}{18 \eta_{air} R} \frac{L}{v_{air}} \left[ \left( \frac{GM_E}{a^2} + \frac{3GM_E R_{eq}^2 J_2}{2a^4}\left(\frac{3\sin^2 u -1}{2}\right) \right) \cos\theta + \frac{q^2 r^2 \cos\xi}{16\pi\varepsilon_0 m(R-r)^2} \right]. (58)$$

Similarly, for elliptical orbits of the same inclinations the ratio takes the form:

$$h_{max} = \frac{(\rho_p - \rho_{air}) d_p^2 C_c}{18 \eta_{air} R} \frac{L}{v_{air}} \left[ \begin{array}{l} \left( \dfrac{GM_E (1+e\cos f)^2}{a^2(1-e^2)} + \dfrac{3GM_E R_{eq}^2 J_2 (1+e\cos f)^4}{2a^4(1-e^2)^4} \right) \cos\theta \\ + \dfrac{q^2 r^2 \cos\xi}{16\pi\varepsilon_0 m(R-r)^2} \end{array} \right], (59)$$

$$h_{max} = \frac{(\rho_p - \rho_{air}) d_p^2 C_c}{18 \eta_{air} R} \frac{L}{v_{air}} \left[ \begin{array}{l} \left( \dfrac{GM_E (1+e\cos f)^2}{a^2(1-e^2)} + \dfrac{3GM_E R_{eq}^2 J_2 (1+e\cos f)^4}{2a^4(1-e^2)^4}\left(\dfrac{3\sin^2 f}{2}-1\right) \right) \cos\theta \\ + \dfrac{q^2 r^2 \cos\xi}{16\pi\varepsilon_0 m(R-r)^2} \end{array} \right], (60)$$

$$h_{max} = \frac{(\rho_p - \rho_{air})d_p^2 C_c}{18\eta_{air}R}\frac{L}{v_{air}}\left[\left(\frac{GM_E(1+e\cos f)^2}{a^2(1-e^2)} + \frac{3GM_E R_{eq}^2 J_2(1+e\cos f)^4}{2a^4(1-e^2)^4}\left(\frac{3\sin^2 f - 1}{2}\right)\right)\cos\theta \right. $$
$$\left. + \frac{q^2 r^2 \cos\xi}{16\pi\varepsilon_0 m(R-r)^2}\right].$$

(61)

## 7. Aerosol Deposition Rate on the Surface of a Planetary Body and in a Spacecraft in Orbit around it.

Following García (2008) we can write the deposition rate $D$ in the following way:

$$D = P_{sed}C_s v_{sed},$$ (62)

where $P_{sed}$ is the sedimentation probability, $v_s$ is its vertical sedimentation velocity, and $C_s$ is the suspending particle concentration. Using Eqs. (2), (7) and (17) we write that on the surface of the Earth the dose takes the form:

$$D = \frac{(\rho_p - \rho_{air})d_p^2 C_c}{18\eta_{air}}\left(\frac{GM_E}{R_{eq}^2\left(1-f'\sin^2\phi\right)^2} - \frac{3GM_E J_2}{2R_{eq}^2\left(1-f'\sin^2\phi\right)^{\frac{3}{2}}}\left(3\sin^2\phi - 1\right) - R_{eq}\left(1-f'\sin^2\phi\right)\omega_E^2\cos^2\phi + \frac{q^2 r^2 \cos\xi}{16\pi\varepsilon_0 m(R-r)^2}\right)\left(1 - e^{-\left\{\frac{(\rho_p - \rho_{air})d_p^2 C_c L_{eq}}{18\eta_{air}Rv_{air}}\left[\left(\frac{GM_E}{R_{eq}^2(1-f'\sin^2\phi)^2} - \frac{3GM_E J_2}{2R_{eq}^2(1-f'\sin^2\phi)^{\frac{3}{2}}}(3\sin^2\phi - 1)\right)\cos\theta + \frac{q^2 r^2 \cos\xi}{16\pi\varepsilon_0 m(R-r)^2}\right]\right\}}\right)C_s.$$ (63)

which for the geocentric latitudes of $\phi = 0°, 45°, 90°$ respectively becomes:

$$D = \frac{(\rho_p - \rho_{air})d_p^2 C_c}{18\eta_{air}}\left(\left(\frac{GM_{eq}}{R_{eq}^2} + \frac{3GM_{eq} J_2}{2R_{eq}^2} - R_{eq}\omega_E^2\right)\cos\theta + \frac{q^2 r^2 \cos\xi}{16\pi\varepsilon_0 m(R-r)^2}\right)\left(1 - e^{-\left\{\frac{(\rho_p - \rho_{air})d_p^2 C_c L_{eq}}{18\eta_{air}Rv_{air}}\left(\left(\frac{GM_{eq}}{R_{eq}^2} + \frac{3GM_{eq} J_2}{2R_{eq}^2} - R_{eq}\omega_E^2\right)\right)\cos\theta + \frac{q^2 r^2 \cos\xi}{16\pi\varepsilon_0 m(R-r)^2}\right\}}\right)C_s,$$ (64)

$$D = \frac{(\rho_p - \rho_{air})d_p^2 C_c}{18\eta_{air}} \left( \left( \frac{GM_{eq}}{R_{eq}^2(1-f')^2} + \frac{3GM_{eq}J_2}{4R_{eq}^2\left(1-\frac{f'}{2}\right)^2} - \frac{1}{2}\left(1-\frac{f'}{2}\right)R_{eq}\omega_E^2 + \frac{q^2 r^2 \cos\xi}{16\pi\varepsilon_0 m(R-r)^2} \right)\cos\theta \right)\left( 1 - e^{-\left( \frac{(\rho_p - \rho_{air})d_p^2 C_c t_{se}}{18\eta_{air}Rv_{se}}\left( \frac{GM_E}{R_{eq}^2(1-f')^2} + \frac{3GM_E J_2}{4R_{eq}^2\left(1-\frac{f'}{2}\right)^2} \right)\cos\theta - \frac{1}{2}\left(1-\frac{f'}{2}\right)R_{eq}\omega_E^2 + \frac{q^2 r^2 \cos\xi}{16\pi\varepsilon_0 m(R-r)^2} \right)} \right) C_s, \quad (65)$$

$$D = \frac{(\rho_p - \rho_{air})d_p^2 C_c}{18\eta_{air}} \left( \left( \frac{GM_E}{R_{eq}^2(1-f')^2} + \frac{3GM_E J_2}{R_{eq}^2(1-f')^2} \right)\cos\theta + \frac{q^2 r^2 \cos\xi}{16\pi\varepsilon_0 m(R-r)^2} \right)\left( 1 - e^{-\left( \frac{(\rho_p - \rho_{air})d_p^2 C_c t_{se}}{18\eta_{air}Rv_{se}}\left( \frac{GM_E}{R_{eq}^2(1-f')^2} + \frac{3GM_E J_2}{R_{eq}^2(1-f')^2} \right)\cos\theta + \frac{q^2 r^2 \cos\xi}{16\pi\varepsilon_0 m(R-r)^2} \right)} \right) C_s. \quad (66)$$

Similarly, in circular and elliptical orbits of inclination $i = 0°$, $45°$, $90°$ we respectively obtain that:

$$D = \frac{(\rho_p - \rho_{air})d_p^2 C_c}{18\eta_{air}} \left( \left( \frac{GM_E}{a^2} + \frac{3GM_E R_{eq}^2 J_2}{2a^4} \right)\cos\theta + \frac{q^2 r^2 \cos\xi}{16\pi\varepsilon_0 m(R-r)^2} \right)\left( 1 - e^{-\left( \frac{(\rho_p - \rho_{air})d_p^2 C_c t_{se}}{18\eta_{air}Rv_{se}}\left( \frac{GM_E}{a^2} + \frac{3GM_E R_{eq}^2 J_2}{2a^4} \right)\cos\theta + \frac{q^2 r^2 \cos\xi}{16\pi\varepsilon_0 m(R-r)^2} \right)} \right) C_s, \quad (67)$$

$$D = \frac{(\rho_p - \rho_{air})d_p^2 C_c}{18\eta_{air}} \left( \left( \frac{GM_E}{a^2} + \frac{3GM_E R_{eq}^2 J_2}{2a^4}\left(\frac{3\sin^2 u}{2} - 1\right) \right)\cos\theta + \frac{q^2 r^2 \cos\xi}{16\pi\varepsilon_0 m(R-r)^2} \right)\left( 1 - e^{-\left( \frac{(\rho_p - \rho_{air})d_p^2 C_c t_{se}}{18\eta_{air}Rv_{se}}\left( \frac{GM_E}{a^2} + \frac{3GM_E R_{eq}^2 J_2}{2a^4}\left(\frac{3\sin^2 u}{2}\right) \right)\cos\theta + \frac{q^2 r^2 \cos\xi}{16\pi\varepsilon_0 m(R-r)^2} \right)} \right) C_s, \quad (68)$$

$$D = \frac{(\rho_p - \rho_{air})d_p^2 C_c}{18\eta_{air}} \left( \left( \frac{GM_E}{a^2} + \frac{3GM_E R_{eq}^2 J_2}{2a^4}\left(\frac{3\sin^2 u - 1}{2}\right) \right)\cos\theta + \frac{q^2 r^2 \cos\xi}{16\pi\varepsilon_0 m(R-r)^2} \right)\left( 1 - e^{-\left( \frac{(\rho_p - \rho_{air})d_p^2 C_c t_{se}}{18\eta_{air}Rv_{se}}\left( \frac{GM_E}{a^2} + \frac{3GM_E R_{eq}^2 J_2}{2a^4}\left(\frac{3\sin^2 u - 1}{2}\right) \right)\cos\theta + \frac{q^2 r^2 \cos\xi}{16\pi\varepsilon_0 m(R-r)^2} \right)} \right) C_s. \quad (69)$$

For elliptical orbits the dose becomes:

$$D = \frac{(\rho_p - \rho_{air})d_p^2 C_c}{18\eta_{air}} \left( \left( \frac{GM_E(1+e\cos f)^2}{a^2(1-e^2)^2} + \frac{3GM_E R_{eq}^2 J_2(1+e\cos f)^4}{2a^4(1-e^2)^4} \right)\cos\theta + \frac{q^2 r^2 \cos\xi}{16\pi\varepsilon_0 m(R-r)^2} \right)\left( 1 - e^{-\left( \frac{(\rho_p - \rho_{air})d_p^2 C_c t_{se}}{18\eta_{air}Rv_{se}}\left( \frac{GM_E(1+e\cos f)^2}{a^2(1-e^2)^2} + \frac{3GM_E R_{eq}^2 J_2(1+e\cos f)^4}{2a^4(1-e^2)^4} \right)\cos\theta + \frac{q^2 r^2 \cos\xi}{16\pi\varepsilon_0 m(R-r)^2} \right)} \right) C_s, \quad (70)$$

$$D = \frac{(\rho_p - \rho_{air})t_z^2 C_c}{18\eta_{air}} \left\{ \left( \frac{GM_E(1 + e\cos f)^2}{a^2(1 - e^2)} \right) + \frac{3GM_E R_M^2 J_2 (1 + e\cos f)^4}{2a^4(1 - e^2)^4} \left( \frac{3\sin^2 f}{2} - 1 \right) \right\} \cos\theta + \frac{q^2 r^2 \cos\xi}{16\pi\varepsilon_0 m(R - r)^2} \left\{ 1 - e^{-\left[ \frac{(\rho_p - \rho_{air})t_z^2 C_c t_{br}}{18\eta_{air} R_{V,m}} \left( \frac{GM_E(1 + e\cos f)^2}{a^2(1 - e^2)} + \frac{3GM_E R_M^2 J_2 (1 + e\cos f)^4}{2a^4(1 - e^2)^4} \left( \frac{3\sin^2 f}{2} - 1 \right) \right) \cos\theta + \frac{q^2 r^2 \cos\xi}{16\pi\varepsilon_0 m(R - r)^2}} \right] \right\} C_s, \quad (71)$$

$$D = \frac{(\rho_p - \rho_{air})t_z^2 C_c}{18\eta_{air}} \left\{ \left( \frac{GM_E(1 + e\cos f)^2}{a^2(1 - e^2)} \right) + \frac{3GM_E R_M^2 J_2 (1 + e\cos f)^4}{2a^4(1 - e^2)^4} \left( \frac{3\sin^2 f}{2} - 1 \right) \right\} \cos\theta + \frac{q^2 r^2 \cos\xi}{16\pi\varepsilon_0 m(R - r)^2} \left\{ 1 - e^{-\left[ \frac{(\rho_p - \rho_{air})t_z^2 C_c t_{br}}{18\eta_{air} R_{V,m}} \left( \frac{GM_E(1 + e\cos f)^2}{a^2(1 - e^2)} + \frac{3GM_E R_M^2 J_2 (1 + e\cos f)^4}{2a^4(1 - e^2)^4} \left( \frac{3\sin^2 f}{2} - 1 \right) \right) \cos\theta + \frac{q^2 r^2 \cos\xi}{16\pi\varepsilon_0 m(R - r)^2}} \right] \right\} C_r,$$

$$\quad (72)$$

## 8. Discussion and numerical results

To proceed with our numerical calculation let us assume the following values for our numerical parameters, i.e., all particles have density $\rho_p = 1300$ kg/m$^3$ (Stuart, 1984), unless range of densities is specified, $\eta_{air} = 1.8 \times 10^{-8}$ kg m$^{-1}$ s$^{-1}$, is the viscosity of the air, $R_{trachea} = d_{trachea}/2 = 25\,\text{mm} = 12.5$ mm (Breatnach et al., 1983), is the radius of the trachea, $\lambda = (0.8\text{-}1.0) \times 10^{-7}$ m is the air mean free path, and $\theta = 2/\pi$ (Landhl, 1963). In fig. 1 we plot of the deposition probability on the surface of the Earth as a function of the particle diameter $d$ and density $\rho$, at geocentric latitude $\phi = 90°$, for particles with diameters and densities in the range $1\mu\text{m} \le d \le 40\mu\text{m}$ and $800$ kg/m$^3 \le \rho \le 2300$ kg/m$^3$, respectively. Similarly, for Mars the Cunningham factor for particles of $1\mu$m in a $CO_2$ atmosphere is given to be $C_M = 15$ (Calle et al. 2011). $J_2 = 0.0010826269$ is the Earth's harmonic coefficient, $J_2 = 0.001964$ (Vallado, 2007), is Mars' harmonic coefficient, $R_M = 3397.2$ km is Mars' radius, $M_M = 6.4191 \times 10^{23}$ kg, is its mass, with a flattering coefficient (Vallado, 2007) and Mars', $\rho_M = 0.02$ kg/m$^3$, is Mars' surface atmospheric density and viscosity and finally $\eta_{air} = 1.47 \times 10^{-5}$ kg m$^{-1}$ s$^{-1}$ (Calle et al. 2011) is Mars' atmospheric viscosity. The deposition probabilities between the poles and the equator for particles of density $\rho_p = 1300$ kg/m$^3$ of various diameters $d$ and residence time $t$ are related according to the relation:

$$P_E\left(\phi = 90^\circ\right) = \left[\frac{1 - e^{-\frac{64137\left(3.2803\times10^{-33} + 7.85859d^3\right)e^{-7.85714 \times 10^6 \phi^2 d}\left(2.8 + \left(8.82 + 5 \times 10^7 d\right)e^{-7.85714 \times 10^6 \phi^2 d}\right)t}{d^2}}}{1 - e^{-\frac{64137\left(3.2803\times10^{-33} + 7.81686d^3\right)e^{-7.85714 \times 10^6 \phi^2 d}\left(2.8 + \left(8.82 + 5 \times 10^7 d\right)e^{-7.85714 \times 10^6 \phi^2 d}\right)t}{d^2}}}\right] P_E\left(\phi = 0^\circ\right). \quad (73)$$

Similarly, on Mars we find that:

$$P_M\left(\phi = 90^\circ\right) = \left[\frac{1 - e^{-\left(3.00435 \frac{3.8743\times10^{-32}}{d^3}\right)d^2 t}}{1 - e^{-\left(2.97823 \frac{3.8743\times10^{-32}}{d^3}\right)d^2 t}}\right] P_M\left(\phi = 0^\circ\right). \quad (74)$$

Next, for the deposition rates on Earth and Mars we respectively obtain:

$$D_E\left(\phi = 90^\circ\right) = 1.00534 \left[\frac{1 - e^{-3.2068 \times 10^9\left(7.85859 \frac{3.2805\times10^{-33}}{d^3}\right)\left(1 + \frac{7\left(1.26 + 0.4e^{-7.85714 \times 10^6 \phi^2 d}\right)}{5\times10^7 d}\right)d^2 t}}{1 - e^{-3.2068 \times 10^9\left(7.85859 \frac{3.2805\times10^{-33}}{d^3}\right)\left(1 + \frac{7\left(1.26 + 0.4e^{-7.85714 \times 10^6 \phi^2 d}\right)}{5\times10^7 d}\right)d^2 t}}\right] D_E\left(\phi = 0^\circ\right), \quad (75)$$

$$D_M\left(\phi = 90^\circ\right) = 1.00001 \left[\frac{1 - e^{-6.2886 \times 10^9\left(2.97823 \frac{3.8743\times10^{-32}}{d^3}\right)d^2 t}}{1 - e^{-6.2886 \times 10^9\left(2.97823 \frac{3.8743\times10^{-32}}{d^3}\right)d^2 t}}\right] D_M\left(\phi = 0^\circ\right). \quad (76)$$

From table 1 we see that particles of a given density and diameter have a greater deposition probability at the poles when compared with that of the equator. On the surface of the Earth, and for a particles of diameter 1, 3, 5 µm, density 1300 kg/m$^3$ and residence times $t = 0.0272$ s we find the following percentage differences for the corresponding deposition probabilities between the poles and the equator to be 0.5%, 0.4%, 0.2% respectively. Similarly, and for the same particles a residence time $t = 0.2$ results to a percentage difference equals to 0.4 %, 0.02 % and 0% between the poles and the equator. From tables 2 and 3 we find that in circular and elliptical orbits around Earth, equatorial orbits i.e. $i = 0°$ result to higher deposition probabilities than the polar orbits, i.e $i = 90°$. In particular for particles of the same diameter, density and residence time $t = 0.0272, 0.2$ s we find the following percentage differences -0.1%, -0.08%, -0.04%, -0.09%, -0.007%, 0% and -0.17%, -0.09%, -0.05% and -0.09%, -0.006%, -0.0001% respectively. Furthermore, on the surface of the Earth and for the same residence times and particles of diameters $d$ =1, 3, 5 μm we find that the deposition probabilities at the poles are related to those at the equator in following way: $P_{\phi=90°} \cong 1.005 P_{\phi=0°}$, $P_{\phi=90°} \cong 1.003 P_{\phi=0°}$ and. $P_{\phi=90°} \cong 1.002 P_{\phi=0°}$. In Earth circular and elliptical orbits of eccentricity $e$ = 0, 0.1 and for particles of the same diameters as the ones given above we find that the deposition probabilities are related in the following way: $P_{\phi=90°} \cong 0.998 P_{\phi=0°}$ or $P_{\phi=90°} \cong 0.999 P_{\phi=0°}$ and $P_{\phi=90°} \cong 0.999 P_{\phi=0°}$ and $P_{\phi=90°} \cong 0.999 P_{\phi=0°}$ or $P_{\phi=90°} \cong 0.999 P_{\phi=0°}$ and $P_{\phi=90°} \cong P_{\phi=0°}$ and for resi-dence time $t = 0.0272$ s and 0.2 s respectively. Similarly, table 4 demon-strates that on the surface of Mars and for the same particle diameters there is an approximately 0.9 % percentage difference between the deposition probabilities at the poles and that at the equator. From tables 5 and 6, we see that for particles of diameters $d$ =1, 3, 5 μm, of the same residence times Martian circular and elliptical orbits at the orbital altitude 300 km with eccentricities i.e. $e$ = 0, 0.1 respectively, result to similar relations for the deposition probabilities i.e. $P_{\phi=90°} = P_{\phi=45°} = P_{\phi=0°}$. This demonstrates that the inclination of the orbit does not really affect the deposition proba-bility. On the other hand if the eccentricity increases then the deposition probability increases as well. On the surface of the Earth/Mars and also in orbit the sedimentation probability becomes zero when angle $\theta$ takes the following value:

$$\theta = \pm\cos^{-1}\left[\left(-\frac{q^2 R^2 x^2 \left(1 + 2f\sin^2\phi + f^2\sin^4\phi\right)\cos\xi}{m\varepsilon_0 (x-R)^2 Q_0}\right)\right] \quad (77)$$

where in the case of the Earth is equal to:

$$Q_0 = 50.2655 GM_E - 37.699 GM_E J_2$$

$$+ 113.097 GM_E J_2 \cos 2\phi + R_E^2 \omega_E^2 \left(\cos^2\phi\left(\begin{array}{c} -50.2655 \\ +f^2\sin^4\phi\left(-150.796 + 50.2655 f\sin^2 f\right) \\ +37.6991 f\sin^2 2\phi \end{array}\right)\right). \quad (78)$$

This angle is independent of the residence time $t$, the particle density $\rho$, and the particle diameter $d$ and depends only on the planetary parameters indicated. Furthermore, it's the same for any geocentric latitude on the Earth and Mars. Numerically for Earth and Mars if the bending angle is equal to $\theta = \pm1.5708°$ the deposition probability/deposition rate becomes equal to zero. On the Earth's surface the deposition probability becomes equal to one if the residence time of a particle of diameter $d_p$ takes the following value:

$$t_{res} = \frac{18\eta R}{d_p^2 (\rho_p - \rho_{air})\left(1 + \frac{2\lambda}{d}\left(\alpha + \beta e^{-\frac{0.55d}{\lambda}}\right)\right)\left(\left(\begin{array}{c} \frac{GM_E}{R_E^2\left(1 - f\sin^2\phi\right)^2} - \frac{3GM_E J_2\left(1 - 3\sin^2\phi\right)}{2R_E^2\left(1 - f\sin^2\phi\right)^2} \\ -R_E\omega_E^2\cos^2\phi\left(1 - f\sin^2\phi\right) \end{array}\right)\cos\theta + \frac{0.0198944 q^2 x^2}{m\varepsilon(x-R)^2}\cos\xi\right)} \quad (79)$$

Similarly, zero deposition probability will imply that $t_{res} = 0$. For particles of diameter $d = 1$ μm, and density $\rho_p = 1300$ kg m$^{-3}$ at the geocentric latitudes of $\phi = 0°$, 45° 90° a deposition probability equal to one will require residence times $t_{res} = 0.764528$, 0.762491, 0.760468 s respectively. Similarly, for a particle with diameter $d_p = 3$ μm of the same density we obtain the following residence times $t_{res} = 0.0943844$, 0.0941329, 0.0938832 s which make the deposition probability equal to one. Thus we see that at the poles the required residence time is smaller than that of the equator. Similarly, on the surface of Mars 1 μm particles result to deposition probability equal to one if the residence time is given by the following expression:

$$t_{res} = \frac{18\eta R}{d_p^2(\rho_p - \rho_{air})\left(1 + \frac{32\lambda}{d}\right)\left(\left(\frac{GM_M}{R_M^2\left(1 - f_M\sin^2\phi\right)^2} - \frac{3GM_M J_2\left(1 - 3\sin^2\phi\right)}{2R_E^2\left(1 - f_M\sin^2\phi\right)^2}\right)\cos\theta + \frac{0.0198944q^2x^2}{m\varepsilon(x-R)^2}\cos\xi\right)} \cdot$$

$$\left(-R_M\omega_M^2\cos^2\phi\left(1 - f_M\sin^2\phi\right)\right)$$

(80)

For dust particles of diameter $d = 1$ μm and mean free path is given by the following expression: $\lambda = 1.6 \times 10^{-5}\left(\frac{T}{p}\right)$ (Read and Lewis, 2004), where $T$ is the absolute temperature in Kelvin, and $p$ is the atmospheric pressure in Pascal. Taking the surface temperature and pressure equals to 210 K and the 610 pa, we find that the mean free path $\lambda = 5.508 \times 10^{-6}$ m = 5.508 μm. Therefore the corresponding residence times become $t_{res} = 3.88128, 3.86430, 3.84751$s respectively. Finally, for an experiment above a spacecraft in an elliptical orbit around Earth the residence time is given by:

$$t_{orbit} \equiv \frac{18R\eta}{d_p^2(\rho_p - \rho_{air})\left[\frac{q^2x^2\cos\xi}{\pi m\varepsilon(x-R)^2} + \frac{8GM_E\cos\theta(1 + e\cos f)^2\left(2a^2\left(1-e^2\right) - 3J_2 R_E^2(1 + e\cos f)^2\left(\sin^2 f\sin^2 i - 1\right)\right)}{a^4\left(1-e^2\right)^4}\right]} \cdot$$

(81)

The electrostatic image force contributes an extra acceleration, that for particles of diameter $d = 1$μm is equal to $1.891 \times 10^{-5}$ μgal, and increases significantly as the particle diameter decreases. For a particle of the same density but with a diameter $d = 0.0017$μm, the image force acceleration is equal to approximately 3849 μgal.

Planetary geophysical parameters can be in principle calculated once a deposition rate can be measured. Using Eq. (63) and expanding the exponen-tial to first order we solve for $J_2$ once the particle concentration $C_{sed}$ and rate of deposition $D$ are known or measured experimentally. Therefore we obtain:

$$J_2 = \frac{(A_0 - C_0)}{B_0} \pm \frac{1}{B_0}\left[H_0 \sec\theta \cos\xi + \frac{18B_0 R\eta_{air}^2 D}{B_0 C_c d_p^2 \eta_{air}\cos\theta\sqrt{RC_{sed}D(\rho^2 - \rho_{air}^2)t}}\right]. \quad (82)$$

Similarly, for a given deposition rate and particle concentration $C_{sed}$, we can obtain solutions for particle diameters $d_p$. Solving Eq. (63) we obtain the following real solution:

$$d_p = 3\sqrt{2\eta_{air}}\left[\frac{RD}{C_{sed}t(\rho - \rho_{air})^2\left(C_c(A_0 - B_0 J_2 - C_0)\cos\theta + C_c H_0 \cos^2\xi\right)}\right]^{1/4}, (83)$$

where:

$$A_0 = \frac{GM_E}{R_{eq}^2\left(1 - f\sin^2\phi\right)^2}, \quad (84)$$

$$B_0 = \frac{3GM_E\left(3\sin^2\phi - 1\right)}{2R_{eq}^2\left(1 - f\sin^2\phi\right)^2}, \quad (85)$$

$$C_0 = R_{eq}\omega_E^2\left(1 - f\sin^2\phi\right)\cos^2\phi \quad , \quad (86)$$

$$H_0 = \frac{q^2 r^2}{16\pi\varepsilon_0 m_p\left(R_{tr} - r\right)^2}. \quad (87)$$

In fig.1 we plot of the deposition probability on the surface of the Earth as a function of the particle diameter $d$ and density $\rho_p$, at the geocentric latitude $\phi = 90°$, for particles with diameters and densities in the range $1\mu m \le d_p \le 5\mu m$ and $1000$ kg/m$^3 \le \rho_p \le 2200$ kg/m$^3$ and residence time $t = 0.0272$ s. The deposition probability increases gradually as the diameter and the particle density increases starting at diameter $d_p \approx 1$ $\mu m$ and $\rho = 1000$ kg/m$^3$. Similarly, in fig. 2 we plot the deposition probability as a function of the particle diameter $d_p$ and geocentric latitude $\phi$ for particles of diameters in the range $1\mu m \le d_p \le 5\mu m$ on the surface of the Earth for the residence time $t = 0.2$ s. We find that the deposition probability gradually increases starting at $d_p \approx 1.5$ μm approaching the higher value plateau but does become not one as the diameter of the particles increase. Therefore, particles in the range of 2 μm and above have a grater deposition probability to be affected by sedimentation, and therefore can end up easily down in the tracheal path. In Fig. 3 we plot of the sedimentation probability on the surface of the Earth as a function of the particle diameter $d_p$ and geocentric latitude $\phi$ for particle diameters and densities in the range $1\mu m \leq d_p \leq 1.0013\mu m$ for residence time $t = 0.2$ s. This figure exhibits the periodic nature of the deposition probability as a function of geocentric latitude when the particle diameter increases by 0.0013 μm, being larger at $\phi = 90°$ and smaller at $\phi = 0°$. In fig. 4, we plot of the deposition probability on the surface of the Earth as a function of the geocentric latitude $\phi$ for particles of diameters 2 μm, density $\rho = 1300$ kg/m$^3$, and residence time $t = 0.2$ s. We find a difference of 0.0006 in the probability of deposition between the latitudes of 90° to 0°. In fig. 5 we plot the deposition probability for an experiment taking place 300 km above the surface of the Earth in a circular polar orbit i.e. $e = 0$ and $i = 90°$, as a function of the particle diameter $d$ and density $\rho$ for particles of diameters in the ranges $1\mu m \leq d \leq 5$ μm and density in the range $1000$ kg/m$^3 \leq \rho \leq 2200$ kg/m$^3$, and residence time $t = 0.2$ s. We find that the sedimentation probability increases significantly for particles of diameters $d > 1.4$ μm, reaching a plateau at approximate diameters $d \geq 3.4$ μm and for the given density range. In fig. 6 we plot the deposition probability in a spacecraft experiment 300 km above the surface of the Earth in a circular equatorial orbit $e = i = 0$, as a function of particle diameter and density and for particles in the diameter in the ranges $1\mu m \leq d \leq 5$ μm and $1000$ kg/m$^3 \leq \rho \leq 2200$ kg/m$^3$ and residence time $t =$ 0.2 s. We find that the deposition probability increases for particles with diameter $d \geq 1.5$ μm reaching gradually a plateau with values less than one at higher densities. In fig. 7 we plot deposition probability as a function of residence time $t$ and particle diameter $d$, for particles of densities 1300 kg m$^{-3}$, in an experiment taking place in a circular orbit of inclination $i = 45°$ We find that a plateau of higher probabilities is reached at $d \approx 3$ μm as the residence time increases. In fig. 8 we plot of the deposition probability for an experiment taking place 300 km above the surface of the Earth in a elliptical polar orbit of eccentricity $e = 0.1$, as a function of the particle diameter $d$, and spacecraft's orbital true anomaly $f$, and residence time $t = 0.2$ s and density $\rho = 1300$ kg/m$^3$. We find that the deposition probability demonstrates a week periodicity in the true anomaly and reaches a plateau of higher probability values for particles of diameter $d \approx 4$ μm. In fig. 9 we plot the deposition probability for an experiment taking place 300 km above the surface of the Earth in a highly eccentric elliptical polar orbit of eccentricity $e = 0.38$, as a function of the particle diameter $d$, and spacecraft's orbital true anomaly $f$, and for the residence a time $t = 0.2$ s and for particles of density $\rho = 1300$ kg/m$^3$. We find that that the deposition probability exhibits a strong periodicity in the true anomaly $f$, between the true anomaly values of $80° \leq f \leq 300°$, reaching a probability plateau at particle diameters $d \geq 4$ μm. Next, in fig. 10 we plot of the deposition probability for an experiment taking place 300 km above the surface of the Earth as a function of the spacecraft's orbital true anomaly $f$ and eccentricity $e$ and for the residence time $t = 0.2$ s for particles of density $\rho = 1300$ kg/m$^3$. We find that the deposition probability demonstrates a periodic effect in the true anomaly effect. The effect demonstrates a plateau effect that is bounded by specific values of the true anomaly namely $20° \leq f \leq 80°$ and $160° \leq f \leq 340°$, and for values of eccentricity in the range $0.6 \leq e \leq 0.8$ and higher.  In fig. 11 we lot of the deposition probability as a function of orbital eccentricity $e$ in an experiment taking place in a spacecraft 300 km above the surface of the Earth in elliptical polar orbits for particles of diameters blue= 1μm, red = 2 μm and density $\rho$ = 1300 kg/m$^3$.  We find that for the same residence time $t$ = 0.2 s the deposition probability increases with particle diameter $d$ and eccentricity $e$.  In particular for particles of diameter 1μm the high probability plateau is reached for eccentricities in the range $0.7 \leq e \leq 0.8$ and higher, and for particles of 2 μm in diameter in the range $0.6 \leq e \leq 0.8$.  Next in fig. 12 we plot of the deposition probability as a function of orbital eccentricity $e$ for particles of diameters $d$ = 3 and 3.5 μm of the same density.  We find that increasing eccentricity increases the deposition probability for the particles of the above diameters.  Larger diameters reach a constant higher probability plateau at smaller eccentricities.  In fig. 13 we plot the deposition probability as a function of orbital inclination $i$ in an experiment taking place in a spacecraft 300 km above the surface of the Earth in a elliptical polar orbit of eccentricity $e$ = 0.1 for particles of the same diameter $d$ = 1.3μm and density $\rho$ = 1300 kg/m$^3$, and corresponding residence times $t$ = 0.0272, 0.2, 0.5 s, respectively from the bottom up.  We find that the orbital inclination does not affect the deposition probability which appears to be constant for the various residence times indicated.  Next, in fig. 14 we plot of the deposition probability on the surface of the Mars as a function of the particle diameter $d$ and density $\rho$, at areocentric latitude $\phi$ = 90°, for particles in the diameter range of $1μm \leq d \leq 5$ μm residence time $t$ = 0.2 s and density 1300 kg/m$^3$.  We find that at the areocentric latitude of 90˚ the deposition probability increases with particle diameter $d$ and density $\rho$ in a much slower way that the corre-sponding Earth one.  In fig. 15 we plot of the deposition probability on the surface of the Mars as a function of the particle diameter $d$ and density $\rho$, at areocentric latitude $\phi = 90°$, for particles in the diameter range of $1\mu m \leq d \leq 5$ $\mu m$, density 1300 kg/m$^3$ and residence time $t = 0.2, 3, 5$ s respectively.  We find that higher particle diameters combined with high densities and higher residence times can easily ensure deposition probabilities that can be equal to one.  For example a residence time of 5 s can easily ensure $P_{dep} = 1.0$ if the density and diameter are high enough.  Fig. 16 we plot the deposition probability as in Fig. 14 but for the areocentric latitude of 45° demonstrates a similar behavior at a lesser degree. In fig. 17, we plot deposition probability versus particle diameter and density and for exactly the same particles with diameter and density ranges of $1\mu m \leq d \leq 5$ $\mu m$ and 1000 kg/m$^3$ $\leq \rho \leq 2200$ kg/m$^3$ respectively.  We find that circular orbits result to a smaller deposition probabilities with a plateau that has a max probability range of approximately half of that of the elliptical orbits.  In fig. 18-19 we plot of the deposition probability on the surface of the Mars as a function of the particle diameter $d$ and density $\rho$, in an experiment taking place in a spacecraft 300 km above the surface of Mars in a elliptical polar orbits of orbital inclination $i = 90°$, and eccentricities $e = 0.1, 0.4$ and residence time $t = 0.2, 3, 5$ s respectively.  We find that the higher inclination considerably affects particle of large diameter/density that fall for higher residence times.  Higher residence times reach a probability plateau at small particle diameters.  In fig. 20 we plot of the deposition probability on the surface of Earth and Mars as a function of particle density $\rho$ for the geocentric and areocentric latitude $\phi = 45°$ and for particles with diameters in the range 1, 2, 3, 4 $\mu m$ for residence time $t = 0.2$ s.  The bottom bundle of straight lines corresponds to Martian deposition probabilities where the upper bundle corresponds to Earth deposition probabilities that are significantly higher when compared to the Martian ones. Fig. 21 gives the deposition rate in an experiment taking place on the surface of Earth as a function particle density $\rho$ and particle concentration $C$ for particles of diameter $d = 1\mu m$ and residence time $t = 0.0272, 0.2$ s. Higher residence times result to higher deposition rates at higher deposition coefficients and densities. For a $t = 0.2$ s deposition plateau it's reached at about $0.10$ particles/m$^2$ s bounded by the density and particle concentration ranges of $2100$ kg/m$^3 \leq \rho \leq 2500$ kg/m$^3$ and $0.6 \leq C \leq 0.8$ respectively. In fig. 22 we plot the information content required in an experiment taking place on the surface of Earth at geocentric latitude $\phi = 90^{\circ}$ as a function particle diameter $d_p$ for particles of density $\rho = 1300$ kg/m$^3$. Similarly, in fig. 23 we plot of the information content required in relation to particles of diameter $d_p$ in a deposition probability experiment on the surface of Mars as a function particle diameter $d_p$. Furthermore, in fig. 24 we plot of the information content required in an experiment taking place on the surface of Earth as a function particle density $\rho_p$. In fig. 24 we plot the information content required in an experiment taking place at various geocentric latitudes on the surface of Earth as a function particle density $\rho$. Finally, in fig. 25 we plot the information content required in an experiment taking place on the surface of Earth as a function the acceleration of gravity $g$ in the range of geocentric latitude $\phi = 0^{\circ}$-$90^{\circ}$.

For residence time $t = 0.2$ s and particles of diameter $d_p = 1\mu m$ on the surface of the Earth the deposition rates numerically become:

$$D_{\phi=0^{\circ}} = 0.262792 C_s, \tag{88}$$

$$D_{\phi=45^{\circ}} = 0.264010 C_s, \tag{89}$$

$$D_{\phi=90^{\circ}} = 0.265228 C_s.$$ (90)

For residence time $t = 0.2$ s and particles of diameter $d_p = 5\mu$m on the surface of the Earth the deposition rates numerically become:

$$D_{\phi=0^{\circ}} = 12.9759 C_s,$$ (91)

$$D_{\phi=45^{\circ}} = 13.0106 C_s,$$ (92)

$$D_{\phi=90^{\circ}} = 13.0452 C_s.$$ (93)

For residence time $t = 0.2$ s and particles of diameter $d_p = 1\mu$m on the surface of Mars the deposition rates numerically become:

$$D_{\phi=0^{\circ}} = 1.74163 \times 10^{-5} C_s,$$ (94)

$$D_{\phi=45^{\circ}} = 1.74164 \times 10^{-5} C_s,$$ (95)

$$D_{\phi=90^{\circ}} = 1.74165 \times 10^{-5} C_s.$$ (96)

For both Earth and Mars and for a particular concentration $C_s$ we find that the deposition rate is higher at the poles comparing to that of the equator. In reference to a spacecraft environment, we say that the atmosphere of spacecraft will contain aerosols (both biological and non-biological) in concentrations which represent the balance between the rates of aerosol particle generation and particle removal by various processes. Physically, the microbial particles while suspended will behave similarly to other particles of various sizes, and at this point there is no need to distinguish about their generation and translocation. In fact, it can be expected that viable microbial particles will physically adhere to nonviable material quite often. During spaceflight there will be the transfer of microbes between crew members. Microbial exchange commonly occurs amongst astronauts. Several bacterial associated diseases were experienced by the crew in Skylab 1. (Taylor et al., 1977) The microbial contamination in the Sky-lab was found to be very high. Staphylococcus aureus and Aspergillus spp have commonly been isolated from the air and surfaces during several space missions. During one mission an increase in the number and spread of fungi and pathogenic streptococci were found in many American Spacecafts (Schuerger 1998). This constitutes an important issue for future extended missions in the Moon or Mars. This consideration could be of particular importance at filter surfaces or in impinger fluids because the nonviable material may serve as substrate permitting the growth of large microbial concentrations on these surfaces. In microgravity conditions it might be necessary to redefine the term "aerosol. " Aerosols consist of particles suspended in gas, and the particles are small enough to stay in suspension for appreciable periods of time. In microgravity, this distinction disappears. However, in the atmosphere of the spacecraft there is still an important difference between the behaviour of large and small particles. The small particles have a relatively ratio of surface-to-mass ratio, and as a result, move through the atmosphere primarily by the influence of viscous drag forces. Large particles, on the other hand, are affected primarily by inertial forces, and are relatively unresponsive to gas flows at moderate velocities. In addition to small particles characteristic of aerosols, the atmosphere of the spacecraft may at times contain relatively large objects from food or equipment. Some of these could lodge in various locations and become potential microbial growth sites. Debris also may enter the equipment used for aerosol control, and interfere in the operation of that equipment. This problem is not encountered in dust collectors in gravitational fields. Coarse objects settle out in the air. In designing aerosol control equipment for the spacecraft environment special attention must be given to means for excluding coarse debris and avoiding interference from this source. Air environments can be important in the dissemination of microbial cells. These are cells, of various sizes that vary in size from a few microns to a fraction of a micron, become suspended in air through a variety of mechanisms. The organisms once suspended remain in air for long periods of time and they can be dispersed over a wide area by atmospheric diffusion. For example a micro-organism with diameter $d$ =1 μm, i.e. the size of some bacteria, settles downward in the air at a rate of only 0.217 m/h or $3.52 \times 10^{-5}$ m/s. Furthermore, small air currents such as those caused by thermal convection are sufficient to propel them in the absence of gravitational attraction. In a spacecraft atmosphere a certain portion of the organisms suspended will remain viable and these organisms can be transported from the interior of space suits and carried by cabin air currents originating from the function of the Environmental Control Systems (ECS) as well as from the motion of the crew. In the absence of gravity, aerosol particles are effectively removed from suspension by contacting a fixed surface to which they are held by mechanical restraint and by chemical or physical interaction. Cabin aerosol particles will move toward the walls and other exposed surfaces at a rate which depends on several factors. The most important driving force will be furnished by the "air" currents which can easily reach random local velocities of several cm per second. Small particles will be carried about by the gas at essentially the same velocities. Statistically, a certain fraction of the particles per unit time will approach a surface closely enough for the particle to be "captured" and held by surface attractive forces. The removal of particles on certain sections of the walls can be increased by increasing the "sticking" probability. This may be accomplished by providing an adhesive surface or by using a dielectric material which readily builds up high electrostatic charges. It might also be possible to utilize the "thermal precipitation" effect, whereby small particles are preferentially deposited on a surface which is maintained colder than surroundings. Further detailed study is needed to determine the advantages of inducing such deposits of aerosols on selected portions of the surface of a space vehicle cabin. The problems of inhibiting microbial growth on the surface and of determining relative collection efficiencies need to be considered in relation to the corresponding problems involved in the use of conventional particle collectors. Inhalation of particles generated as a result of thermal degradation from fire or smoke, as may occur on spacecraft, is of major health concern to space-faring countries. Knowledge of lung airflow and particle transport under different gravity environments is required to addresses this concern by providing information on particle deposition. Similarly deposition of inhaled lunar dust particles will be very important in the case of future moon missions. The extent of the inflammatory response in the lung will depend on where the Lunar and Mars dust particles are deposited. During 1g deposition in the more central airways will reduce the transport of the fine particles in the periphery of the lung. On the other hand the fractional gravity of the Moon and Mars will deposit the inhaled particles in more peripheral lung regions. Moon's less gravity will also result in a smaller sedimentation rate. As a result fine dust particles will deposit in the alveolar lung region, exacerbating the potential for the lung damage. Lunar dust is known to have a large surface area (i.e., it is porous), and a substantial portion is in the respirable range. The surface of the lunar dust particles is known to be chemically activated by processes ongoing at the surface of the moon.

An alternative framework for the study of gravitational effects in biology, and in particular the interaction of gravity and the respiratory system is the introduction of information. The introduction of this different approach discussed below requires a detailed description of Shannon linear model of communication, and details on the theory of information and cybernetics which can be found in several references such as (Shannon 1948; Shannon et al 1959; Berger 1971). In order to formalize the gravity role in the sedimentation process and to help the reader follow the rationale behind it, we provide some interpretations proper to Shannon information model. Information theory, which was originally developed for use in telecommunications, has in recent years been increasingly applied to analyzing biological signaling pathways to answer relative biological questions (Porter et. al., 2012). For Shannon, anything is a source of information (in our case, the respiratory system of the human organism) if it has a number of alternative states (in our case, states of gravitational interaction) that might be realized on a particular occasion and the method of encoding is based on the presupposition. It is necessary to process, or encode information from a source (in our case the encoder is the gravitational field and the encoding is done through the gravitational acceleration formula which codifies all changes of gravitational interaction) before transmitting it through a given channel. According to Shannon information presupposes the existence of a signal probability, where the signal can be described via a stochastic distribution function (Shannon, 1948). Furthermore, this signal propagates through communicational channels (Shannon, 1948). With reference to the communication channel we say, that this constitutes a composite channel (particles, environment, respiratory system as well as the physiology of the system), where a way of transportation of the signal is the sedimentation (itself), among which the basic regulatory parameters are gravity itself as well as the rest of the variables.

Even though there is still no final definition of the term information, according to Stonier (1990) "information exists" and regulates various parameters of the system that they are impossible to be described in a deterministic model, as well as its not possible to quantify via a stochastic model (e.g. in the case of the respiratory system these factors could be the structural form, variation in lung capacities, coordination of muscles for controlling breathing, exchange gases with the circulatory system, etc.). The net effect of these parameters can be modeled in the theory of information. According to Shannon (Shannon, 1948) information is transferred with an allowable degree of noise. Shannon allows the introduction of noise in the interaction of a biological organism and environment, where noise accounts for biological environment variation (Burton and Iglesias 2007, Tkaik, and Walczak, 2011, Porter et. al., 2012).

In this paper we do not quantify the effect of noise according to Shannon's rate distortion theory, but we study how information or the self-information of the system can be reflected and codified via gravity in the natural phenomenon of sedimentation, taking into account that the respiratory system is itself a decoder of the information transmitted and contained in the system of the human body. A decoder is a device which does the reverse operation of an encoder, undoing the encoding so that the original information can be retrieved. In our case the decoder is the respiratory system. Next, according to Shannon, we consider the probability of sedimentation $P_{sed}$ to be a random variable and remind the reader that according to Shannon the amount of the information content, is given in (Mac-Kay, 2005):

$$I(x = a_i) = -\log_2 \left[ P(x = a_i) \right], \tag{97}$$

is a sensible measure of the information content of the outcome $x = a_i$. Therefore, let us consider that the sedimentation is the random variable and that the amount of information content associated with the sedimentation of dust particles in the respiratory system is:

$$I(P_{sed}) = -\log_2(P_{sed}) = -\log_2\left[1 - e^{-\left(\frac{(\rho_p - \rho_{air})d_p^2 C_c}{18\eta_{air}R}tg\cos\theta\right)}\right] \text{ bits,} \quad (98)$$

from which we can also obtain that the dependence of the acceleration of gravity itself on the amount of information content is given by:

$$g(I) = \frac{q^2 r^2 \cos\xi}{16\pi a n_p (r-R)^2} + \frac{18\eta R \sec\theta}{C_c(\rho_p - \rho_{air})d_p^2 t}\log\left(1 - 2^{-I}\right). \quad (99)$$

Next, $g = g(I)$ since sedimentation is taking place but also since this sedimentation connects to the acceleration of gravity. The fact that the two variables are correlated is manifested on the phenomenon of sedimentation. Therefore, the spatial-temporal conditions (gravity, time, etc.) serve as substrate for the correlation of gravity $g$ and information $I$. In the case of a dust particle that has diameter $d = 1\mu m$ and density $\rho = 1300$ kg/m$^3$ on the surface of the Earth, we obtain that the information related to the sedimentation has a dependence on the acceleration gravity $g$ that is given by the equation below:

$$I = -1.4427\ln\left(1 - e^{-0.00037726\left(8.8742 \times 10^{-14} + 0.8041\,g\right)}\right) \quad (100)$$

From Eq. (85) we that if the acceleration of gravity $g$ changes the probability of sedimentation and therefore changes the information contained in the random variable, as it is decoded by the pulmonary system. Thus we say that the pulmonary system is a receiver and a decoder of information. Therefore, the environment is the sender as it sends information on the human body, in other words the biophysics of sedimentation equals to the a communication channel. This communication channel can be designed and based on parameters such as gravity to change. Therefore gravity in our case controls how to transmit the information signal. The pulmonary system is the biological decoder in which information is received and as primordial information it interacts with the structural information of the system, producing a new kind of information which could be interpreted as a kind of meta-information information of the interaction. This can be thought as a way in which gravity enters the cognitive processes of the system (processing of information) in the cybernetic sense. The "Shannon" approach to sedimentation is a new different approach that is related to the main approach adopted in the paper through the probability of sedimentation. This is a first step towards an approach that aims in a deeper understanding and mathematical and structural description of information processing in human organisms in variable gravitational background.

**Table 1** Sedimentation probability geocentric latitude effect for an experiment taking place on surface of the Earth and for particle of various diameters and density $\rho = 1300$ kg/m$^3$.

| Geocentric Latitude $\phi$ [°] | Particle diameter $d$[μm] | Sedimentation Probability $t = 0.0272$ s | Sedimentation Probability $t = 0.2$ s |
|---|---|---|---|
| 0 | | 0.077080 | 0.445561 |
| 45 | 1.0 | 0.077277 | 0.446434 |
| 90 | | 0.077475 | 0.447304 |
| 0 | 3.0 | 0.477816 | 0.991583 |
| 45 | | 0.478722 | 0.991689 |
| 90 | | 0.479624 | 0.991795 |
| 0 | | 0.828767 | 0.999998 |
| 45 | 5.0 | 0.829572 | 0.999998 |
| 90 | | 0.830372 | 0.999998 |

**Table 2** Sedimentation probability inclination and eccentricity effect for an experiment taking place in a circular orbit above the Earth for particle of various diameters and density $\rho = 1300 \text{ kg/m}^3$.

| Orbital Inclination $i$[°] and Eccentricity $e = 0$ | Particle diameter $d$[µm] | Sedimentation Probability $t = 0.0272$ s | Sedimentation Probability $t = 0.2$s |
|---|---|---|---|
| 0 | | 0.0707827 | 0.4171340 |
| 45 | 1.0 | 0.0707096 | 0.4167970 |
| 90 | | 0.0707020 | 0.4167620 |
| 0 | | 0.4482460 | 0.9873800 |
| 45 | 3.0 | 0.4478950 | 0.9873210 |
| 90 | | 0.4478580 | 0.9873140 |
| 0 | | 0.8011340 | 0.9999930 |
| 45 | 5.0 | 0.8007890 | 0.9999930 |
| 90 | | 0.8007590 | 0.9999930 |

**Table 3** Sedimentation probability inclination and eccentricity effect for an experiment taking place in an elliptical orbit above the Earth for particle of various diameters and density $\rho = 1300 \text{ kg/m}^3$ .

| Orbital Inclination $i$[°] and Eccentricity $e = 0.1$ | Particle diameter $d$[µm] | Sedimentation Probability $t = 0.0272$ s | Sedimentation Probability $t = 0.2$ s |
|---|---|---|---|
| 0 | | 0.07216880 | 0.4324970 |
| 45 | 1.0 | 0.07209280 | 0.4321500 |
| 90 | | 0.07208500 | 0.4231140 |
| 0 | 3.0 | 0.45487800 | 0.9884530 |
| 45 | | 0.45481700 | 0.9883970 |
| 90 | | 0.45447900 | 0.9883910 |
| 0 | | 0.80755900 | 0.9999950 |
| 45 | 5.0 | 0.80721200 | 0.9999940 |
| 90 | | 0.80717600 | 0.9999940 |

**Table 4** Sedimentation probability geocentric latitude effect for an experiment taking place on surface of the Mars and for particles of diameter $d$ and density $\rho = 1300$ kg/m³.

| Geocentric Latitude $\phi$ [°] | Particle diameter $d$[μm] | Sedimentation Probability $t = 0.0272$ s | Sedimentation Probability $t = 0.2$ s |
|---|---|---|---|
| 0 | | 0.000509296 | 0.00373877 |
| 45 | 1.0 | 0.000511533 | 0.00375516 |
| 90 | | 0.000513765 | 0.00377152 |

**Table 5** Sedimentation probability geocentric latitude effect for an experiment taking place in a Mars circular orbit and for particles of diameter $d$ and density $\rho = 1300$ kg/m³.

| Orbital Inclination $i$[°] and Eccentricity $e = 0$ | Particle diameter $d$[μm] | Sedimentation Probability $t = 0.0272$ s | Sedimentation Probability $t = 0.2$ s |
|---|---|---|---|
| 0 | | 0.000431798 | 0.00317063 |
| 45 | 1.0 | 0.000431798 | 0.00317063 |
| 90 | | 0.000431798 | 0.00317063 |

**Table 6** Sedimentation probability geocentric latitude effect for an experiment taking place in a Mars elliptical orbit and for particles of diameter $d$ and density $\rho = 1300$ kg/m³.

| Orbital Inclination $i$[°] and Eccentricity $e = 0.1$ | Particle diameter $d$[μm] | Sedimentation Probability $t = 0.0272$ s | Sedimentation Probability $t = 0.2$ s |
|---|---|---|---|
| 0 | | 0.00173888 | 0.0127155 |
| 45 | 1.0 | 0.00173888 | 0.0127155 |
| 90 | | 0.00173888 | 0.0127155 |

**Table 7** Sedimentation probability geocentric latitude effect for an experiment taking place in a Mars elliptical orbit of high orbital eccentricity $e = 0.3$ for particles of diameter and density $\rho = 1300$ kg/m$^3$

| Orbital Inclination $i[°]$ and Eccentricity $e = 0.3$ | Particle diameter $d[\mu\text{m}]$ | Sedimentation Probability $t = 0.0272$ s | Sedimentation Probability $t = 0.2$ s |
|---|---|---|---|
| 0 | | 0.00173888 | 0.0190125 |
| 45 | 1.0 | 0.00173888 | 0.0190125 |
| 90 | | 0.00173888 | 0.0190125 |

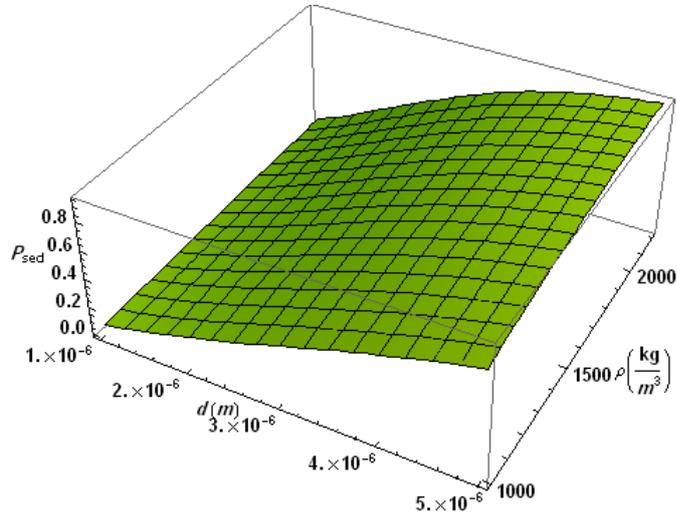

**Fig. 1** Plot of the deposition probability on the surface of the Earth as a function of the particle diameter $d$ and density $\rho$, at geocentric latitude $\phi = 90°$, for particles of diameters $d$ in the ranges $1\mu\text{m} \leq d \leq 5$ µm and $1000$ kg/m$^3 \leq \rho \leq 2200$ kg/m$^3$, and residence time $t = 0.0272$ s.

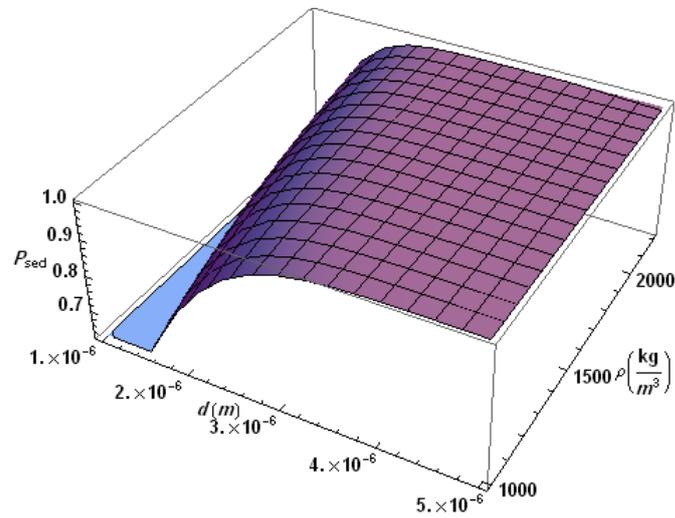

**Fig. 2** Plot of the deposition probability on the surface of the Earth as a function of the particle diameter $d$ and density $\rho$, at geocentric latitude $\phi = 90°$, for particles of diameter $d$ in the ranges $1\mu\text{m} \le d \le 5$ µm and density $\rho$ in the range $1000$ kg/m$^3 \le \rho \le 2200$ kg/m$^3$, and residence time $t = 0.2$ s

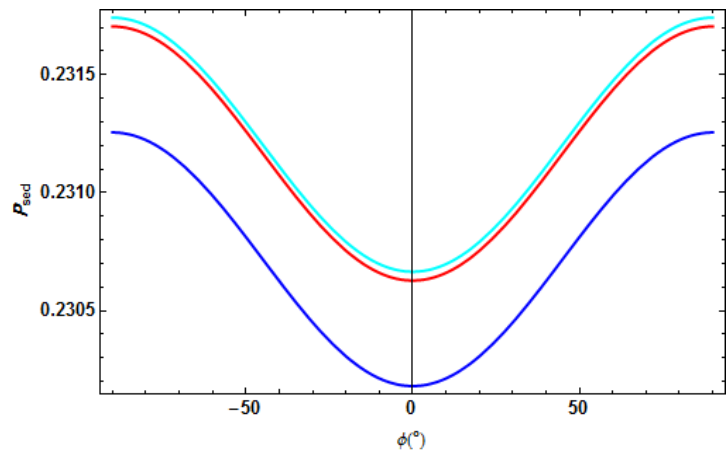

**Fig. 3** Plot of the deposition probability on the surface of the Earth as a function of geocentric latitude $\phi$ for particles of various diameters in the range $1\mu\text{m} \le d \le 1.0013$ µm of density $\rho = 1300$ kg/m$^3$, and residence time $t = 0.2$ s.

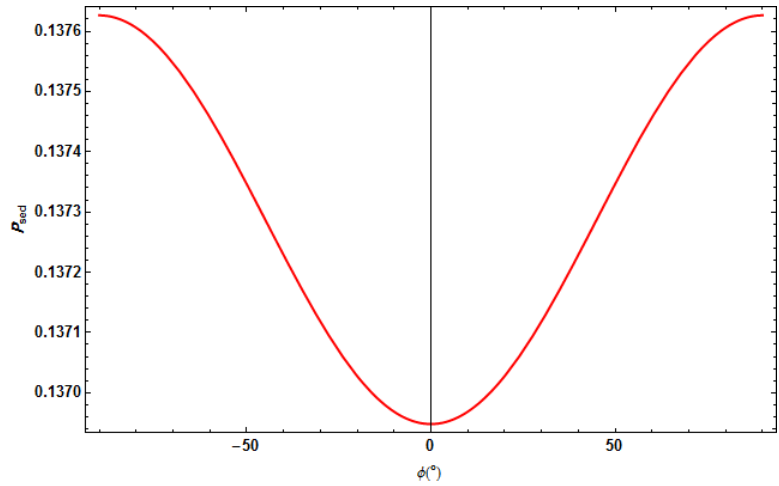

**Fig. 4** Plot of the deposition probability on the surface of the Earth as a function and geocentric latitude $\phi$ for particles of diameters 2 µm, density $\rho = 1300$ kg/m$^3$, and residence time $t = 0.2$ s.

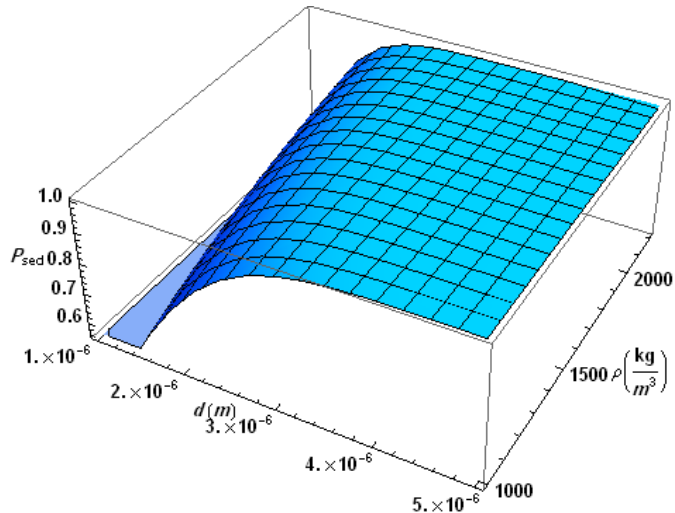

**Fig. 5** Plot of the deposition probability in an experiment taking place in spacecraft in a circular polar orbit $e = 0$ and $i = 90^\circ$, 300 km above the surface of the Earth as a function of the particle diameter $d$ and density $\rho$ for particles of diameters $d$ in the ranges $1\,\mu\text{m} \leq d \leq 5$ µm and density in the range 1000 kg/m$^3 \leq \rho \leq 2200$ kg/m$^3$, and for residence time $t = 0.2$ s.

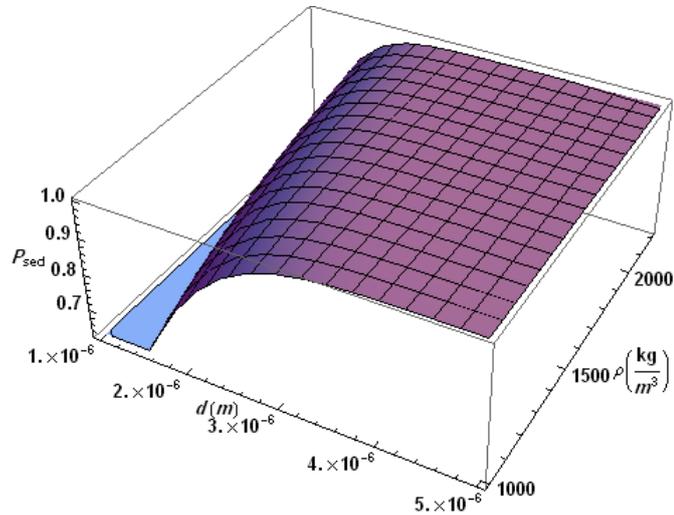

**Fig.6** Plot of the deposition probability in an experiment taking place in spacecraft in a circular equatorial orbit $e = 0$ and $i = 0°$, 300 km above the surface of the Earth as a function of the particle diameter $d$ and density $\rho$, for particles of diameters $d$ in the ranges $1\,\mu m \leq d \leq 5\,\mu m$ and density in the range $1000\,kg/m^3 \leq \rho \leq 2200\,kg/m^3$, and for residence time $t = 0.2$ s.

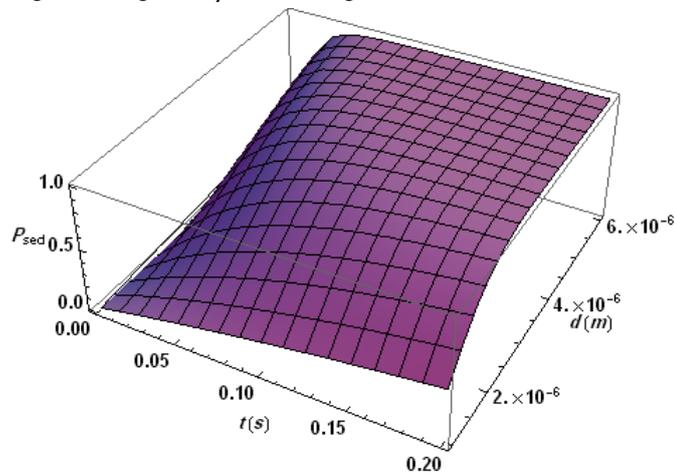

**Fig.7** Plot of the deposition probability in an experiment taking place in spacecraft in a circular orbit $e = 0$ and $i = 45°$, 300 km above the surface of the Earth as a function of the particle residence time $t$, and diameter $d$, particle density used $\rho = 1300$ kg/m³.

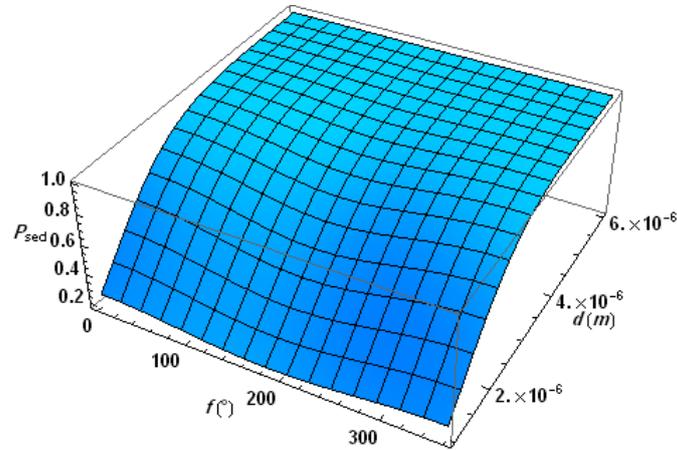

**Fig. 8** Plot of the deposition probability for an experiment taking place 300 km above the surface of the Earth in a elliptical polar orbit, as a function of the particle diameter $d$, and spacecraft's orbital true anomaly $f$, eccentricity used $e = 0.1$, residence time $t = 0.2$ s and particle density $\rho = 1300$ kg/m$^3$.

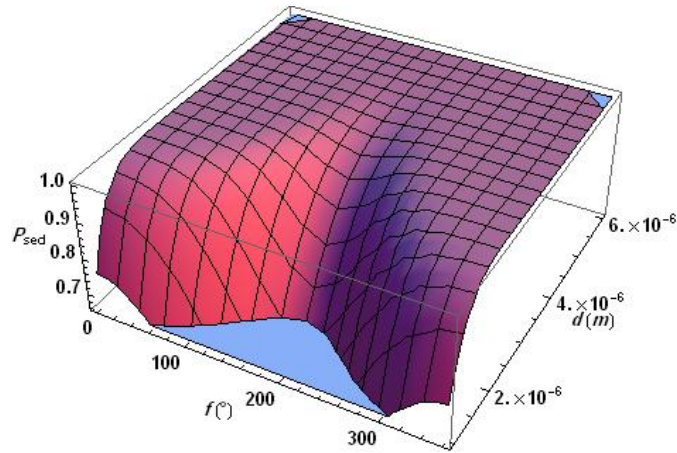

**Fig. 9** Plot of the deposition probability for an experiment taking place 300 km above the surface of the Earth in a elliptical polar orbit, as a function of the particle diameter $d$, and spacecraft's orbital true anomaly $f$, orbital eccentricity used $e = 0.38$, residence time $t = 0.2$ s for particle density $\rho = 1300$ kg/m$^3$.

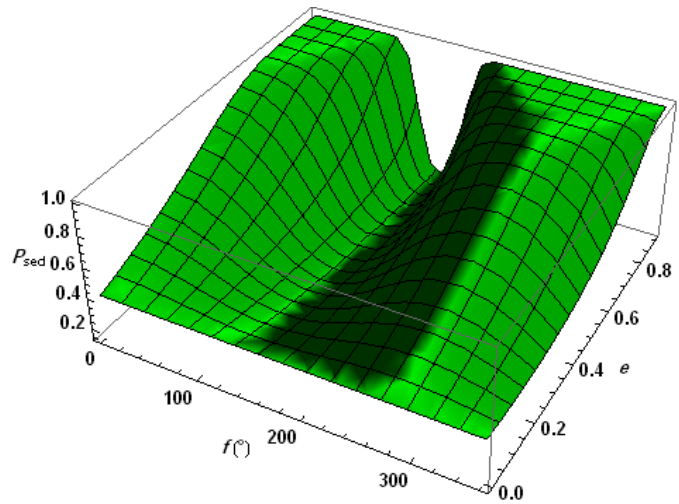

**Fig. 10** Plot of the deposition probability for an experiment taking place 300 km above the surface of the Earth in a elliptical polar orbit, as a function of the spacecraft's orbital true anomaly $f$ and eccentricity $e$ for residence time $t = 0.2$ s and density $\rho = 1300$ kg/m$^3$.

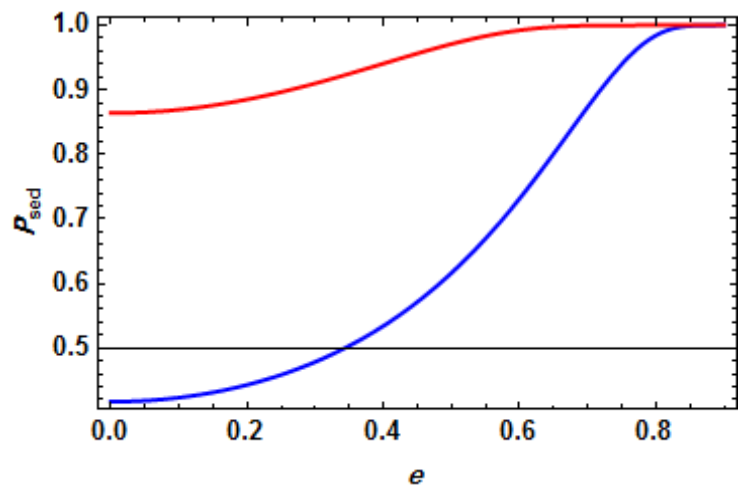

**Fig. 11** Plot of the deposition probability as a function of orbital eccentricity $e$ in an experiment taking place in a spacecraft 300 km above the surface of the Earth in a elliptical polar orbit for particles of diameters blue= 1μm red = 2 μm and density $\rho = 1300$ kg/m$^3$.

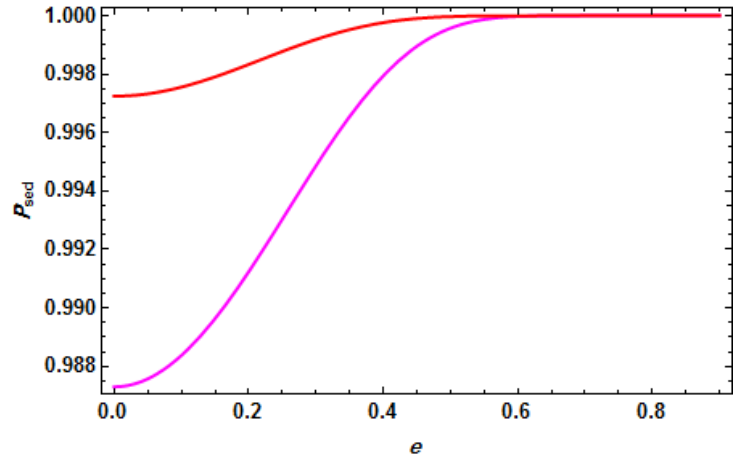

**Fig. 12** Plot of the deposition probability as a function of orbital eccentricity $e$ in an experiment taking place in a spacecraft 300 km above the surface of the Earth in a elliptical polar for particles of diameters magenta= 3μm and red = 3.5 μm and density $\rho = 1300$ kg/m³.

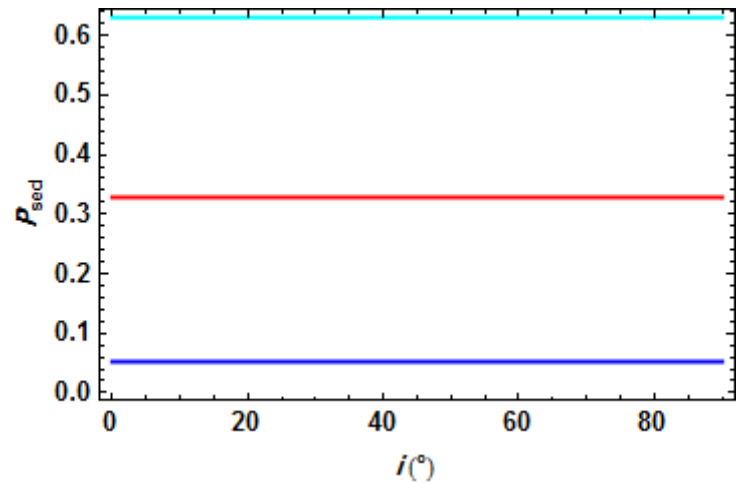

**Fig. 13** Plot of the deposition probability as a function of orbital inclination $i$ in an experiment taking place in a spacecraft 300 km above the surface of the Earth in a elliptical polar orbit of eccentricity $e = 0.1$ for particles of the same diameter $d = 1.3$μm and density $\rho = 1300$ kg/m³, and corresponding residence times $t = 0.0272, 0.2, 0.5$ s, respectively from the bottom up $d = 1.3$μm.

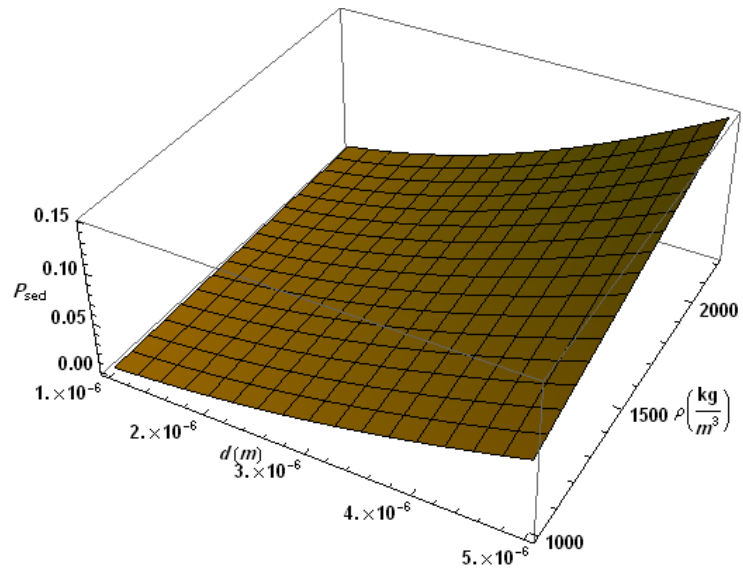

**Fig. 14** Plot of the deposition probability on the surface of the Mars as a function of the particle diameter $d$ and density $\rho$, at areocentric latitude $\phi = 90°$, for particles in the diameter/density range of $1\,\mu m \leq d \leq 5\,\mu m$ and $1000\ kg/m^3 \leq \rho \leq 2200\ kg/m^3$ and residence time $t = 0.2$ s respectively.

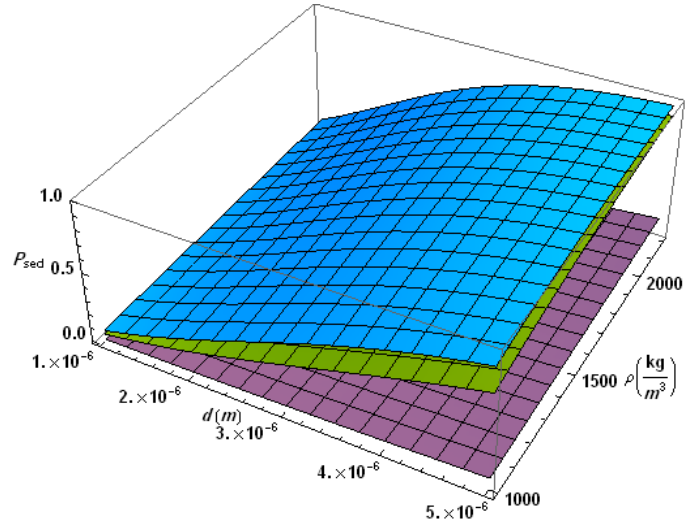

**Fig. 15** Plot of the deposition probability on the surface of the Mars as a function of the particle diameter $d$ and density $\rho$, at areocentric latitude $\phi = 90°$, for particles in the diameter/density range of $1\,\mu m \leq d \leq 5\,\mu m$ and $1000\,kg/m^3 \leq \rho \leq 2200\,kg/m^3$ and residence times $t = 0.2, 3, 5$ s respectively.

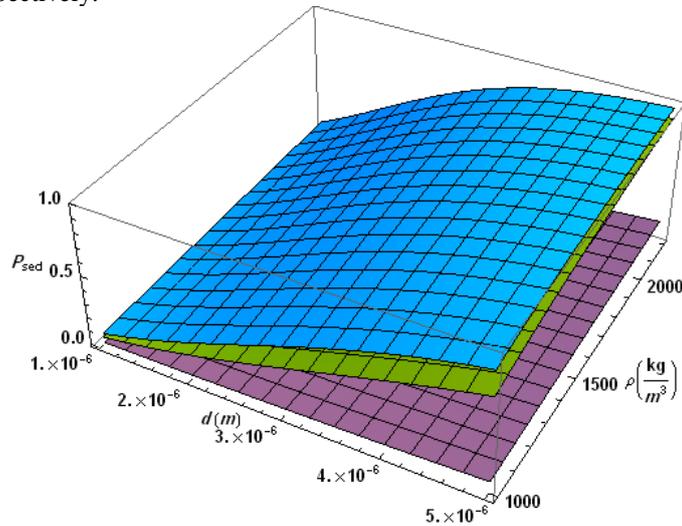

**Fig. 16** Plot of the deposition probability on the surface of the Mars as a function of the particle diameter $d$ and density $\rho$, at areocentric latitude $\phi = 45°$ for particles for particles in the diameter/density range of $1\,\mu m \leq d \leq 5\,\mu m$ and $1000\,kg/m^3 \leq \rho \leq 2200\,kg/m^3$ and residence times $t = 0.2, 3, 5$ s respectively.

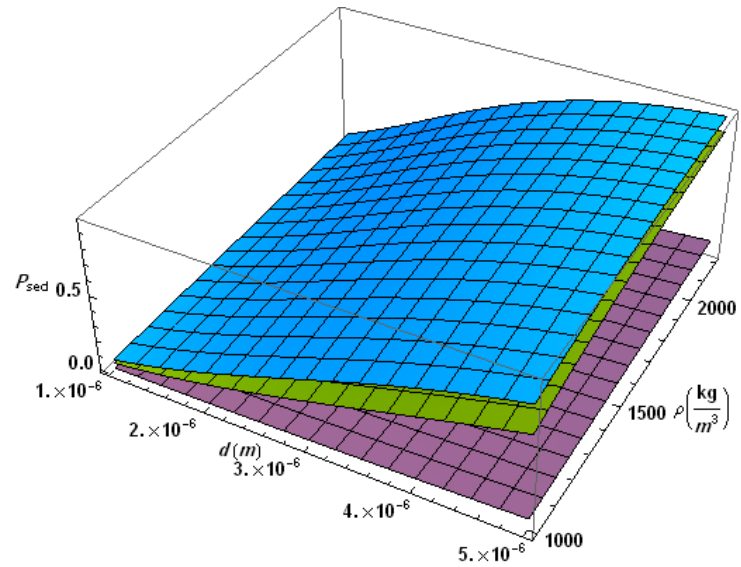

**Fig. 17** Plot of the deposition probability on the surface of the Mars as a function of the particle diameter $d$ and density $\rho$, in an experiment taking place in a spacecraft 300 km above the surface of Mars in a circular polar orbit of orbital inclination $i = 90°$, eccentricity $e = 0$ and residence time $t = 0.2, 3, 5$ s respectively, and for particles in the in the diameter/density range of $1\,\mu\text{m} \leq d \leq 5\,\mu\text{m}$ and $1000\ \text{kg/m}^3 \leq \rho \leq 2200\ \text{kg/m}^3$.

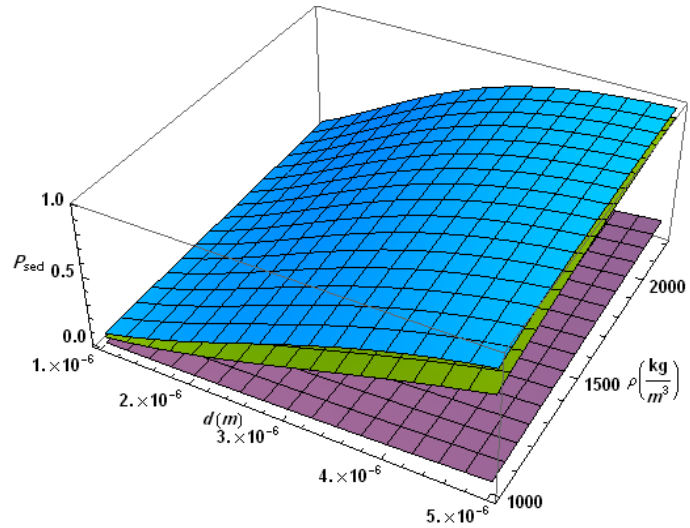

**Fig. 18** Plot of the deposition probability on the surface of the Mars as a function of the particle diameter $d$ and density $\rho$, in an experiment taking place in a spacecraft 300 km above the surface of Mars in a elliptical polar orbit of orbital inclination $i = 90°$, eccentricity $e = 0.1$ and residence time $t = 0.2$, 3, 5 s respectively, respectively, and for particles in the diameter/density range of 1 μm ≤ $d$ ≤ 5 μm and 1000 kg/m³ ≤ $\rho$ ≤ 2200 kg/m³.

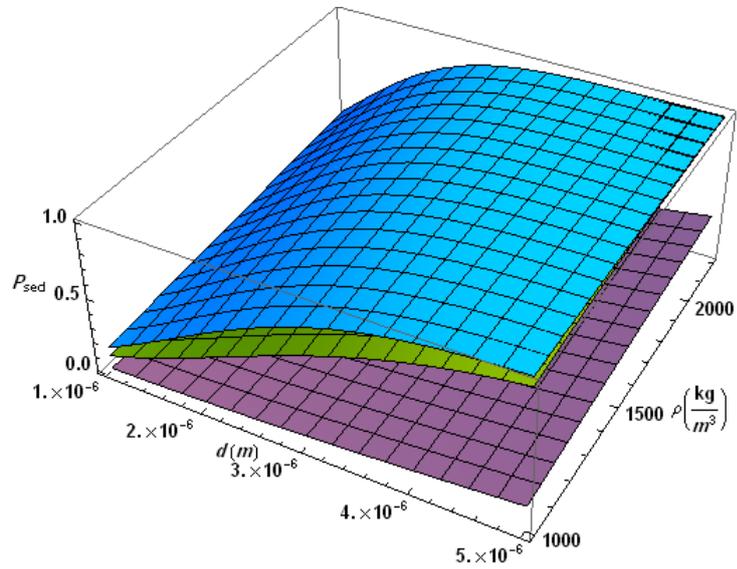

**Fig. 19** Plot of the deposition probability on the surface of the Mars as a function of the particle diameter $d$ and density $\rho$, in an experiment taking place in a spacecraft 300 km above the surface of Mars in a elliptical polar orbit of orbital inclination $i = 90°$, eccentricity $e = 0.4$ and residence time $t = 0.2$, 3, 5 s respectively, and for particles in the in the diameter/density range of $1\,\mu m \leq d \leq 5\,\mu m$ and 1000 kg/m$^3 \leq \rho \leq 2200$ kg/m$^3$.

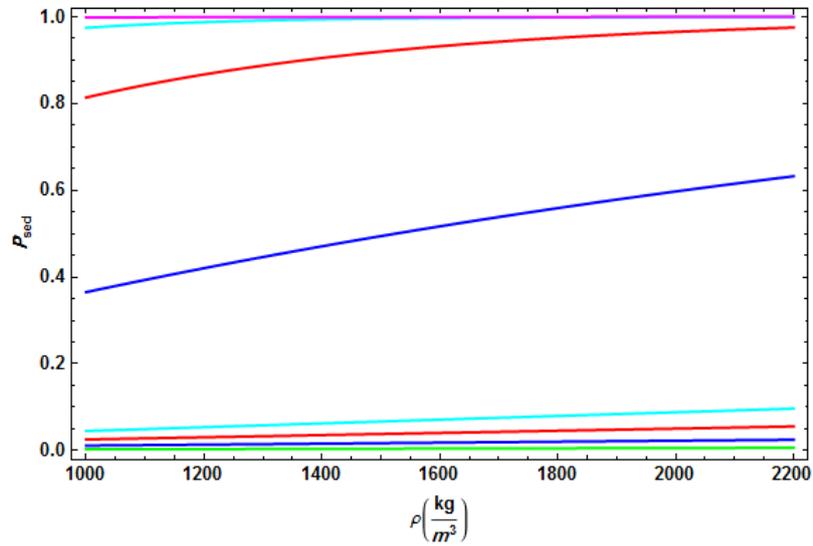

**Fig. 20** Plot of the deposition probability on the surface of Earth and Mars as a function of particle density $\rho$ for the geocentric and areocentric latitude $\phi = 45°$ for particles of diameters in the range $1\mu m \leq d \leq 4$ μm for residence time $t = 0.2$ s. The bottom bundle of straight lines corresponds to Mars.

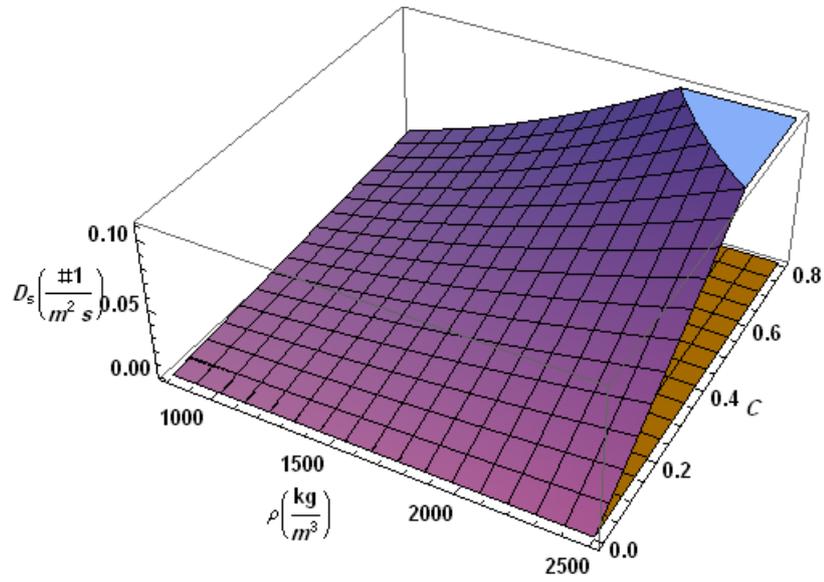

**Fig. 21** Plot of the deposition rate in an experiment taking place on the surface of Earth as a function particle density $\rho$ and particle concentration $C$ for particles of diameter $d = 1\mu m$ and residence times $t = 0.0272,\ 0.2$ s correspondingly.

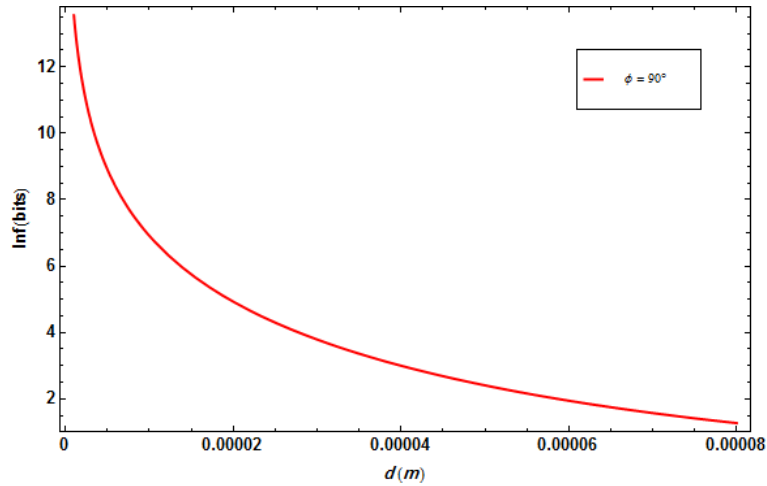

**Fig. 22** Plot of the information content required in an experiment taking place on the surface of Earth at geocentric latitude $\phi = 90°$ as a function particle diameter $d$ for particles of density $\rho = 1300$ kg/m$^3$.

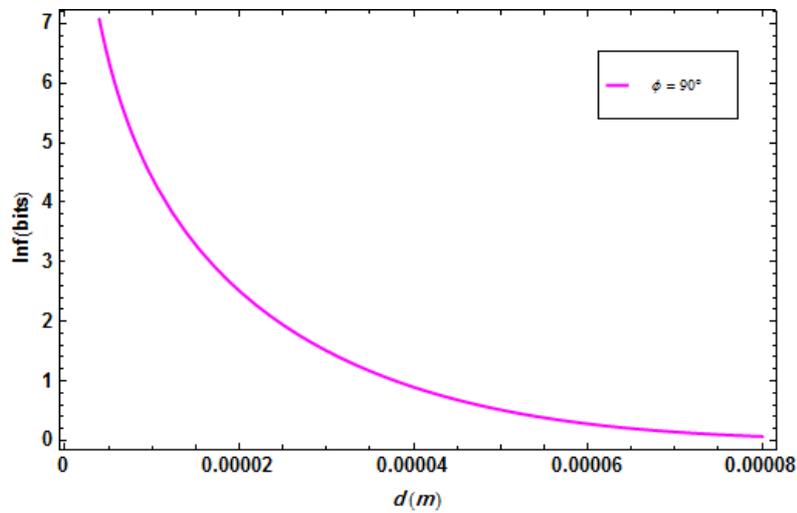

**Fig. 23** Plot of the information content required in relation to particles of diameter $d$ in the sedimentation probability experiment surface of Mars as a function particle diameter $d$ particles of density $\rho = 1300$ kg/m$^3$.

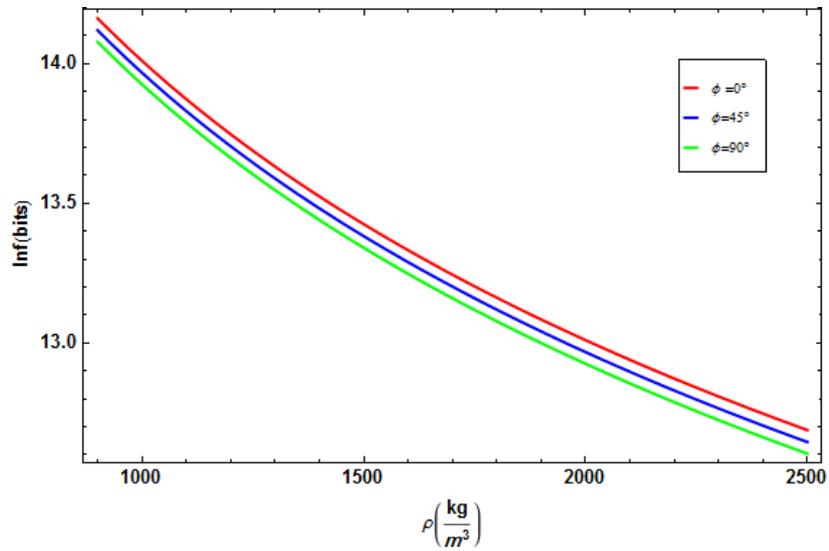

**Fig. 24** Plot of the information content required in an experiment taking place at various geocentric latitudes on the surface of Earth as a function particle density $\rho$.

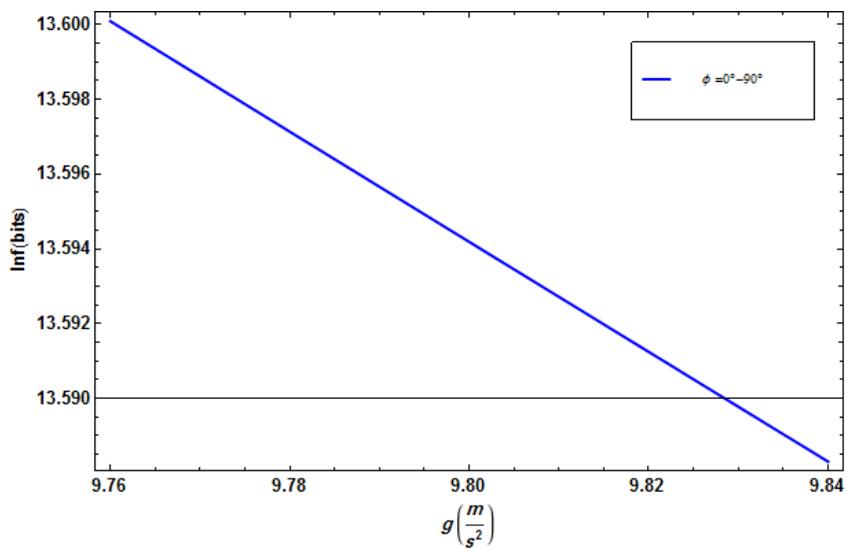

**Fig. 25** Plot of the information content required in an experiment taking place on the surface of Earth as a function the acceleration of gravity $g$ in the range of geocentric latitude $\phi = 0^{\circ}$-$90^{\circ}$.

## 9. Conclusions

We have studied the effect of the acceleration of gravity on the deposition probability due to sedimentation. In doing so, on the surface of the Earth and Mars we have corrected the acceleration of gravity for the zonal harmonic coefficients of both planets as well as for their rotation. On the surface of the Earth/Mars, have found that deposition probability due to sedimentation at the poles is higher that that at the equator. For 1 μm particles of density 1300 kg/m$^3$ with residence time of 0.2 s exhibit a 0.4 % percentage difference, where on Mars the corresponding percentage difference is 0.9%. When orbiting around a planet equatorial orbits result to higher deposition probabilities than the polar ones. In particular for a spacecraft orbiting at 300 km above the Earth in a circular orbit we find that 1μm particles with corresponding residence times of 0.0272 and 0.2 s exhibit -0.1% and -0.09% percentage difference in the deposition probabilities between polar and equatorial orbits. Similarly elliptical orbits of eccentricity $e = 0.1$ exhibit a percentage difference equal to -0.17% and -0.09 % respectively. On the surface of Mars and for the residence time of 0.2 s the deposition probability of 1 μm particles demonstrates an approximately 0.9% percentage difference between poles and the equator. Similarly, and for the same particles there is a 0% difference between circular/elliptical polar and equatorial orbits. We have also found that high eccentricity elliptical orbits result to higher deposition probabilities. Finally, as an alternative framework for the study of interaction and the effect of gravity in biology the term information according to Shannon has been used, and in particular gravity and respiratory system. This can be thought as a way in which gravity enters the cognitive processes of the system (processing of information) in a bio-gravitational cybernetic sense, introducing the possi-bility of a multidisciplinary approach of the study of the effects of gravity on humans. This may be important also because it could open interesting perspectives to project new kinds of missions having both biological and fundamental physics and bio-engineering goals.

**Acknowledgements:** The authors of this paper would like to thank two anonymous reviewers for their encouraging review of this paper.